\documentclass{sig-alternate}  % SIGMOD REJECT
\usepackage{graphicx} 
\graphicspath{ {./pics/}{.} }  % the spacing is required.
\DeclareGraphicsExtensions{.pdf,.png,.jpg}  % order preference for images of same name.
\usepackage{balance}
\usepackage{wrapfig}
\usepackage{amsmath}
\usepackage{amssymb}
\usepackage{algorithm}
\usepackage{algorithmicx}
\usepackage{algpseudocode}
\usepackage{booktabs}
\usepackage{enumerate}
\usepackage{comment}
\usepackage{cite}
\usepackage{url}

\usepackage[T1]{fontenc}

\usepackage{xspace}

\usepackage{color}
\title{Exploiting Opportunistic Physical Design in Large-scale Data Analytics}

\numberofauthors{1}

\newcommand{\expandauth}{{\global\advance\auwidth by 40pt}}
\newcommand{\contractauth}{{\global\advance\auwidth by -55pt}}
\newcommand{\hakanname}{{\fontencoding{T1}Hakan Hac{\i}g\"{u}m\"{u}\c{s}}}

\author{
 \alignauthor Jeff LeFevre\raisebox{1.5mm}{$^{+*}$}~~~~~Jagan Sankaranarayanan\raisebox{1mm}{$^{*}$}~~~~~\hakanname\raisebox{1.5mm}{$^{*}$} \\ Junichi Tatemura\raisebox{1.5mm}{$^{*}$}~~~~Neoklis Polyzotis\raisebox{1.5mm}{$^{+}$}~~~~Michael J. Carey\raisebox{1.5mm}{$^{\%}$}\\
 \fontsize{9}{9}\selectfont\sffamily\upshape{\raisebox{1.5mm}{$^*$}NEC Labs America, Cupertino, CA~~~\raisebox{1.5mm}{$^+$}University of California, Santa Cruz~~~\raisebox{1.5mm}{$^\%$}University of California, Irvine}\\
 \vspace{0.1in}
 \fontsize{7}{7}\selectfont\ttfamily\upshape{\{jlefevre,alkis\}@cs.ucsc.edu, \{jagan,hakan,tatemura\}@nec-labs.com, mjcarey@ics.uci.edu}\
 \vspace{0.1in}
}

\newcommand{\onefig}[3]{
 \begin{figure}[!h]
 \vspace{-0.1in}
  \centering % SJ: Apparently, \centering is more space efficient that \begin{center}
    \includegraphics[width=#3]{#1}
    \vspace{-0.1in}
    \caption{\label{#1}#2}
    \vspace{-0.07in}
 \end{figure}
}

\newcommand{\onefigtop}[3]{
 \begin{figure}[!t]
 \vspace{-0.1in}
  \centering % SJ: Apparently, \centering is more space efficient that \begin{center}
    \includegraphics[width=#3]{#1}
    \vspace{-0.1in}
    \caption{\label{#1}#2}
    \vspace{-0.07in}
 \end{figure}
}

\newcommand{\twofigrow}[4]{
 \begin{figure*}[t!]
 \vspace{-0.1in}
 \centering
    \includegraphics[width=#4]{#1}%
    \includegraphics[width=#4]{#2}\\
    \vspace{-0.2in}
    {\tiny{\hspace{0.3in}{\sf (a)}\hspace{#4}\hspace{-0.1in}{\sf (b)}}}
    \vspace{-0.13in}
    \caption{\label{#1}{#3}}
 \vspace{-0.1in}
 \end{figure*}
}

\newcommand{\twofig}[4]{
 \begin{figure}[h!]
 \vspace{-0.2in}
 \centering
    \includegraphics[width=#4]{#1}%
    \includegraphics[width=#4]{#2}\\
    \vspace{-0.07in}
    {\tiny{\hspace{0.3in}{\sf (a)}\hspace{#4}\hspace{-0.1in}{\sf (b)}}}
    \vspace{-0.13in}
    \caption{\label{#1}{#3}}
 \vspace{-0.2in}
 \end{figure}
}

\newcommand{\threefig}[5]{
 \begin{figure*}[t!]
 \vspace{-0.1in}
\begin{minipage}{\linewidth}
\centering
    \includegraphics[width=#5]{#1}%
    \includegraphics[width=#5]{#2}%
    \includegraphics[width=#5]{#3}\\
    \vspace{-0.3in}
    {\tiny{\hspace{0.3in}{\sf (a)}\hspace{#5}\hspace{-0.1in}{\sf (b)}\hspace{#5}\hspace{-0.1in}{\sf (c)}}}
\end{minipage}
 \vspace{-0.15in}
    \caption{\label{#1}#4}
 \vspace{-0.10in}
 \end{figure*}
}

\renewcommand{\algorithmicforall}{\textbf{for each}}
\renewcommand{\algorithmicindent}{8pt}
\newcommand{\Line}{\vspace{-0.08in}\hrulefill\vspace{-0.04in}}
\algblock[Init]{Init}{EndInit}

\newcommand{\WF}{\ensuremath{\mbox{\sc Bf\-Rewrite}}\xspace}
\newcommand{\VF}{\ensuremath{\mbox{\sc View\-Finder}}\xspace}
\newcommand{\minCost}{\ensuremath{\mbox{\sc best\-Plan\-Cost}}\xspace}
\newcommand{\merge}{\ensuremath{\mbox{\sc Merge}}}
\newcommand{\minPlumb}{\ensuremath{\mbox{\sc best\-Plan}}}

\newcommand{\PropagateBestRewrites}{\ensuremath{\mbox{\sc Prop\-Best\-Rewrite}}}
\newcommand{\RefineTarget}{\ensuremath{\mbox{\sc Refine\-Target}}}

\newcommand{\cost}{\ensuremath{\mbox{\sc Cost}}}
\newcommand{\ocost}{\ensuremath{\mbox{\sc OptCost}}\xspace}
\newcommand{\qlet}{\ensuremath{\leftarrow}}

\newcommand{\RE}{\ensuremath{\mbox{\sc Rewrite\-Enum}}\xspace}
\newcommand{\PQ}{\ensuremath{\mbox{PQ}}}
\newcommand{\FindNextMinTarget}{\ensuremath{\mbox{\sc Find\-Next\-Min\-Target}}}

\renewcommand{\Init}{\ensuremath{\mbox{\sc Init}}}
\newcommand{\Peek}{\ensuremath{\mbox{\sc Peek}}}
\newcommand{\Refine}{\ensuremath{\mbox{\sc Refine}}}
\newcommand{\node}[1]{\ensuremath{\mbox{\sc node}_{#1}}}
\newcommand{\complete}{\ensuremath{\mbox{\sc GuessComplete}}}

\newtheorem{definition}{Definition}
\newtheorem{theorem}{Theorem}
\newtheorem{example}{Example}

\newcommand\fakesc[1]{{\relsize{-1}\MakeUppercase{#1}}}

\begin{document}
\maketitle

%\blfootnote{\raisebox{2mm}{{\tiny $\#$}}Work done while author was at NEC Labs America}\\

%\input{Introduction-Old} % Original Version
%\input{Introduction-Hakan} % Hakan Version
%\input{Introduction-merged}  % Merged Version
%\input{Introduction-rewrite7} % Jeff version per Alkis comments
%\input{Introduction-rewrite8} % Alkis version 

\vspace{-0.6in}
\begin{abstract}
Big data analytical systems, such as MapReduce, perform aggressive 
materialization of intermediate job results in order to support fault 
tolerance. 
When jobs correspond to exploratory queries submitted by data analysts, 
these materializations yield a large set of materialized views that typically 
capture common computation among successive queries from the same analyst, or 
even across queries of different analysts who test similar hypotheses. 
We propose to treat these views as an opportunistic physical design and use 
them for the purpose of query optimization. 
We develop a novel query-rewrite algorithm that addresses the two main 
challenges in this context: 
how to reason about views that contain UDFs (a common feature in big data 
analytics), and how to search the large space of rewrites.
%To do this, we first develop a UDF model that captures a common class of UDFs
%  and enables effective reuse of previous results.
To do this, we first develop a semantic UDF model that captures an important 
class of UDFs for big data analysis:  MapReduce UDFs containing arbitrary code.  
The model enables effective reuse of previous results generated by UDFs.
We then present a rewrite algorithm, inspired by nearest-neighbor searches in 
metric spaces, that provably finds the minimum-cost rewrite under certain 
 assumptions.
An extensive experimental study  on real-world datasets using our prototype based on Hive shows that our 
approach results in dramatic performance improvements for complex big data
analysis queries --- reducing total execution time over 60\% on average and 
up to an order of magnitude. 
\end{abstract}

\section{Introduction}
Data analysts have the crucial task of analyzing the ever increasing volume of 
data that modern organizations collect in order to produce actionable insights. 
As expected, this type of analysis on big data is highly exploratory in nature and involves
 an iterative process: the data analyst starts with an initial query over the 
data, examines the results, then reformulates the query and may even bring in 
additional data sources, and so on. 
Typically, these queries involve sophisticated, domain-specific operations 
that are linked to the type of data  and the purpose of the analysis, e.g., 
performing sentiment analysis over tweets or computing network influence. 
Because a query is often revised multiple times in this scenario, there can 
be significant overlap between queries.
There is an opportunity to speed up these explorations by reusing previous
query results either from the same analyst or from different analysts 
performing a related task.

MapReduce (MR) has become a de-facto tool for this type of analysis.
It offers scalability to large datasets, easy incorporation of new data
sources, the ability to query right away without defining a schema up front, 
and extensibility through user-defined functions (UDFs).
Analyst queries are often written in a declarative query language, e.g.,
HiveQL or PigLatin, which are automatically translated to a set of MR jobs.
Each MR job involves the materialization of intermediate results (the 
 output of mappers, the input of reducers and the output of reducers) for 
 the purpose of failure recovery. A typical Hive or Pig query will spawn a 
 multi-stage job that will involve several such materializations. 
We refer to these execution artifacts as 
\textit{opportunistic materialized views}.

We propose to treat these views as an \emph{opportunistic physical design} and 
to use them to rewrite queries.
The opportunistic nature of our technique has several nice properties: the
materialized views are generated as a by-product of query execution, i.e., 
without additional overhead; the set of views is naturally tailored to the 
current workload; and, given that large-scale analysis systems typically 
execute a large number of queries, it follows that there will be an equally 
large number of materialized views and hence a good 
chance of finding a good rewrite for a new query. 
Our results indicate the savings in query execution time can be dramatic: 
a rewrite can reduce execution time by up to an order of magnitude.

Rewriting a query using views in the context of MR involves a unique 
combination of technical challenges. 
%First, the queries and views almost certainly use UDFs containing arbitrary 
%user code, thus query rewriting requires some \emph{semantic} understanding of 
%the UDFs.
First, the queries and views almost certainly contain UDFs, thus query rewriting 
requires some \emph{semantic} understanding of UDFs.
These MR UDFs for big data analysis are composed of arbitrary user-code and  
may involve a sequence of MR jobs.
Second, any query rewriting algorithm that can utilize UDFs now has to contend 
with a potentially large number of operators since any UDF can be included 
in the rewriting process.
Third, there can be a large search space of views to consider for rewriting 
due to the large number of materialized views in the opportunistic physical 
design, since they are almost free to retain (storage permitting).

Recent methods to reuse MR computations such as ReStore~\cite{Elgh2012}
and MRShare~\cite{Li2011} lack any semantic understanding of execution 
artifacts and can only reuse/share cached results when execution plans are 
syntactically identical.
%We strongly believe that any truly effective solution will have to address 
% the query rewrite problem in its full generality.
We strongly believe that any truly effective solution will have to a 
incorporate a deeper semantic understanding of cached results and 
``look into'' the UDFs as well.

\textbf{Contributions.} In this paper we present a novel query-rewrite 
algorithm that targets the scenario of opportunistic materialized views 
in an MR system with queries that contain UDFs.
We propose a UDF model that has a limited semantic understanding 
of UDFs, yet enables effective reuse of previous results.
Our rewrite algorithm employs techniques inspired by spatial databases
 (specifically, nearest-neighbor searches in metric spaces~\cite{Hjal03}) in
 order to provide a cost-based incremental enumeration of the huge space of
 candidate rewrites, generating the optimal rewrite in an efficient manner.
Specifically, our contributions can be summarized as follows:
%- Gray-box model for UDFs.
%- Rewrite algorithm.
%- Experimental study.
\begin{itemize}
  \item A gray-box UDF model that is simple but expressive enough to capture 
  a large class of MR UDFs that includes many common analysis tasks.
  The UDF model further provides a  quick way to compute a lower-bound on the cost 
  of a potential rewrite given just the query and view definitions.   
   We provide the model and the types of UDFs it admits in 
   Sections~\ref{sec-udf-model}--\ref{sec-using-model}.
  
  \item A rewriting algorithm that uses the lower-bound to 
  (a) gradually explode the space of rewrites as needed, and 
  (b) only attempts a rewrite for those views with good potential to produce 
  a low-cost rewrite.
  We show that  the algorithm produces the optimal rewrite as well as finds 
  this rewrite in a work-efficient manner, under certain assumptions.
  %The work-efficient property means the algorithm will not consider a 
  %candidate view unless it has the potential to form the optimal rewrite.
  We describe this further in 
  Sections~\ref{sec-best-first-workflow-rewrite}--\ref{sec-viewfinder}.
  
  \item An experimental evaluation showing that our methods provide execution time 
  improvements of up to an order of magnitude using real-world data and 
  realistic complex queries containing UDFs.
  The execution time savings of our method are due to 
  %reusing opportunistic views from long-running computations. This results in 
  moving much less data and avoiding the high expense of re-reading data from 
  raw logs when possible.
  We describe this further in Section~\ref{sec-experimental-evaluation}.  
   
\end{itemize}

\section{Preliminaries}\label{sec-preliminaries}
Here we present the architecture of our system
and briefly describe its components and how they interact, 
followed by our notations and problem definition.

\subsection{System Architecture}\label{sec-system-architecture}

Figure~\ref{fig-odyssey-system-diagram} provides a high level
overview of our system and its components.
Our system is built on top of Hive, and queries are written in HiveQL.
Queries are posed directly over log data stored in HDFS.
In Hive, MapReduce UDFs are given by the user as a series of Map or Reduce 
jobs containing arbitrary user code expressed in a supported language such 
as Java, Perl, Python, etc.
To reduce execution cost, our system automatically rewrites queries based on 
the existing views.
A query execution plan in Hive consists of a series of MR jobs, 
and each MR job materializes its output to HDFS.
As Hive lacks a mature query optimizer and cannot cost UDFs, 
we implemented an optimizer based on the cost model 
from~\cite{Nyki2010} and extended it to cost UDFs, as described
later in Section~\ref{sec-costing-udfs}.

During query execution, all by-products of query processing 
(i.e., the intermediate materializations) are retained as opportunistic 
materialized views.
These views are stored in the system (space permitting) as the opportunistic 
physical design.

The materialized view metadata store contains information about the
materialized views currently in the system such as the view definitions 
and standard data statistics used in query optimization.
For each view stored, we collect statistics by running a lightweight Map 
job that samples the view's data.
This constitutes a small overhead, but as we show experimentally in 
Section~\ref{sec-experimental-evaluation}, this time is a small fraction of 
query execution time.

The rewriter, presented in Section~\ref{sec-best-first-workflow-rewrite}, 
uses the materialize view metadata store to rewrite queries based on the 
existing views.
%In order to let the rewriter communicate with the optimizer of the execution 
%engine, 
To facilitate this, our optimizer generates plans with two types of annotations on each plan node: 
(1) the logical expression of its computation (Section~\ref{sec-udf-example}) and 
(2) the estimated execution cost (Section~\ref{sec-costing-udfs}).

\onefigtop{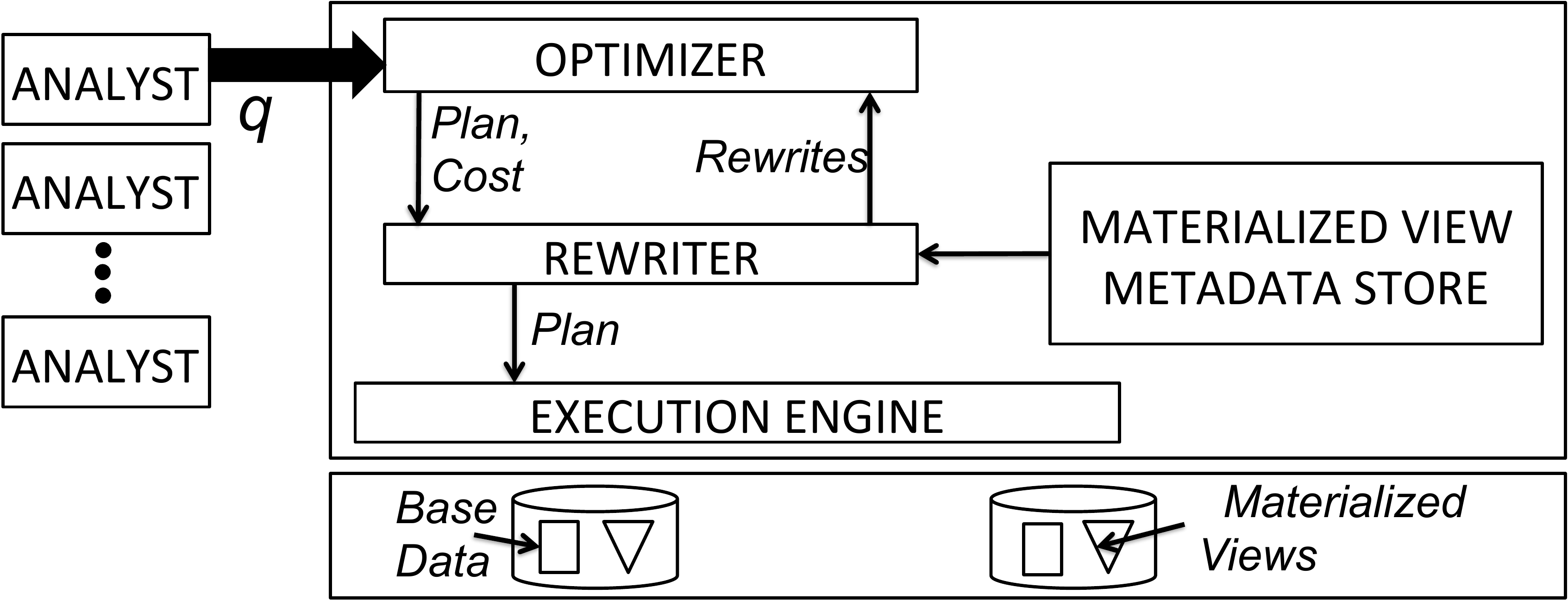}
{System diagram showing control flows.}{3.25in}

The rewriter uses the logical expression in the annotation when searching for
rewrites for each node in the plan. The expression consists of
relational operators or UDFs. For each rewrite found
during the search, the rewriter utilizes the optimizer to obtain
an estimated cost for the rewritten plan.
% plan and estimated cost.

\subsection{Notations}\label{sec-preliminaries-definitions}

$W$ denotes a plan generated by the query optimizer, which is
represented as a DAG containing $n$ nodes, ordered topologically.
Each node represents an MR job.
We denote the $i^{th}$ node of $W$ as \node{i}, $i \in [1,n]$.
The plan has a single sink that computes the result of the query; under the 
topological order assumption the sink is \node{n}.
%$W_i$ is a sub-graph that contains all nodes up to and including \node{i}.
$W_i$ is a sub-graph of $W$ containing \node{i} and all of its ancestor nodes.
We refer to $W_i$ as one of the rewritable \emph{targets} of plan $W$.
%A property of $W_i$ is that it represents a materialization point in plan $W$.
As is standard in Hive, the output of each job is materialized to disk.
Hence, a property of $W_i$ is that it represents a materialization point 
in $W$, and in this way, materializations are free except for statistics 
collection.
An outgoing edge from \node{k} to \node{i} represents data flow 
from $k$ to $i$.
$V$ is the set of all opportunistic materialized views (MVs) in the system.
% which includes the base data.

%%% The function  $\cost(W_i)$ takes as input $W_i$ and returns its estimated 
%%% cost as provided by the query optimizer.
We use $\cost(\node{i})$ to denote the cost of executing the MR job at \node{i},
as estimated by the query optimizer.
Similarly, $\cost(W_i)$ denotes the estimated cost of running the sub-plan 
rooted at $W_i$, which is computed as 
$\cost(W_i) = \sum_{\forall \node{k} \in W_i} \cost(\node{k})$.
% We make two assumptions about the optimizer.
% First that it can provide a cost estimate for relational operators, and 
 %second that it can provide a cost estimate for a UDF.
%%%\cost(\node{i}) is the cost of the MR job at \node{i}, given its input data.
%%%\cost($W_i$) $ = \sum_{\forall \node{k} \in W_i} \cost(\node{k})$.

%We use $r_i$ to denote an equivalent rewrite of target $W_i$ using views 
% in $V$.
We use $r_i$ to denote an equivalent rewrite of target $W_i$ \textsl{iff} 
$r_i$ uses only views in $V$ as input and produces an identical output to 
$W_i$, for the same database instance $D$.
A rewrite $r^*$ represents the minimum cost rewrite of $W$ 
(i.e., target $W_n$).
%In this work we only consider equivalent rewrites.

\subsection{Problem Definition}
Given these basic definitions, we introduce the problem we solve in this paper.

\textbf{Problem Statement.}
\textit{Given a plan $W$ for an input query $q$, and a set of materialized 
views $V$, find the minimum cost rewrite $r^*$ of $W$.}\looseness=-1  

Our rewrite algorithm considers views in $V$ during the search for $r^*$.  
Since some views may contain UDFs, for the rewriter to utilize those views 
during its search, some understanding of UDFs is required.
Next we will describe our UDF model and then present our rewrite algorithm 
that solves this problem.

\section{UDF Model}\label{sec-udf-model}
Since big data queries frequently include UDFs, in order to reuse previous 
computation in our system effectively we desire a way to model MR UDFs 
semantically.
If the system has no semantic understanding of the UDFs, then the opportunities
for reuse will be limited --- essentially the system will only be able to exploit 
cached results when one query applies the exact same UDF to the exact same input
as a previous query.  
However, to the extent that we are able to ``look into'' the UDFs and 
understand their semantics, there will be more possibilities for reusing 
previous results.
In this section we propose a UDF model that allows a deeper semantic 
understanding of MR UDFs.
Our model is general enough to capture a large class of UDFs that 
includes classifiers, NLP operations (e.g., taggers, sentiment), text 
processors, social network (e.g., network influence, centrality) and spatial 
(e.g., nearest restaurant) operators.  
Of course, we do not require the developer to restrict herself to this model; 
rather, to the extent a query uses UDFs that follow this model, the 
opportunities for reuse will be increased.

\subsection{Modeling a UDF}

\onefig{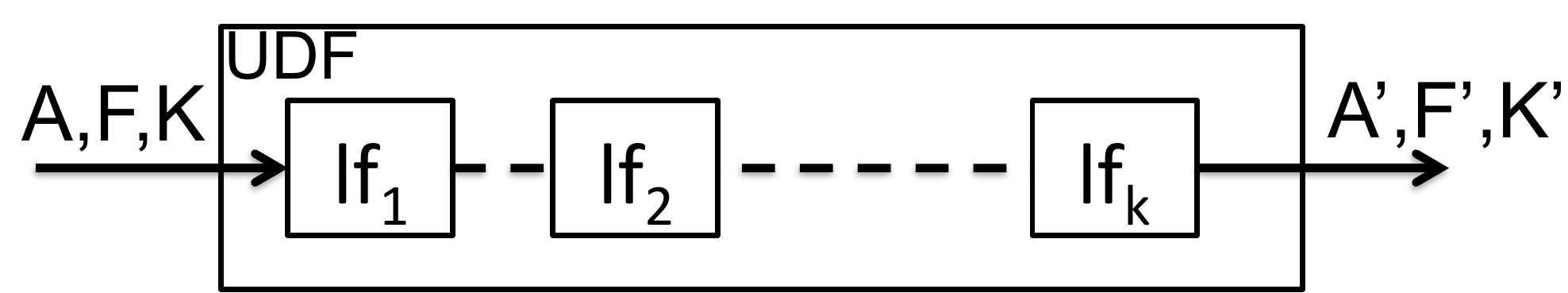}
{A UDF composed of local functions (\emph{lf}$_1$, \emph{lf}$_2$, 
$\cdots$, \emph{lf}$_k$), showing the end-to-end transformation of input to 
output.}{2in}

We propose a model for UDFs that allows the system to capture a UDF as  
a {\em composition} of {\em local functions} as shown in 
Figure~\ref{fig-composition-udf-local-function}, where each local function 
represents a map or reduce task. 
The nature of the MR framework is that map-reduce functions are stateless and
only operate on subsets of the input, i.e., a single tuple or a single group of tuples.
Hence, we refer to these map-reduce functions as {\em local functions}.
A local function can only 
perform a combination of 
% These operations correspond exactly 
the following three types of operations 
performed by \textit{map} and \textit{reduce} tasks.
%We restrict a local function to perform the following operations.
\begin{enumerate}
\item Discard or add attributes, where an added attribute and its values may 
be determined by arbitrary user code
\item Discard tuples by applying filters, where the filter predicates may be 
performed by arbitrary user code
\item Perform grouping of tuples on a common key, where the grouping operation 
may be performed by arbitrary user code
\end{enumerate}
% Notice also that the semantics of these local functions and filters
% are opaque unlike a white-box model.

The end-to-end transformation of a UDF is obtained by
composing the operations performed by each local function \emph{lf} in the UDF.
Our model captures the fine-grain dependencies between the input and output tuples
in the following way.

The UDF input is modeled as $(A, F, K)$ where $A$ is the set of attributes, 
$F$ is set of filters previously applied to the input, 
and $K$ is the current grouping of the input, which captures the keys of the data.
The output is modeled as $(A', F', K')$ with the same semantics. 
Our model describes a UDF as the transformation from $(A,F,K)$ to $(A',F',K')$ 
as performed by a composition of local functions using operation types 
(1) (2) (3) above.
Figure~\ref{fig-composition-udf-local-function} shows how to semantically model
a UDF that takes any arbitrary input represented as $A,F,K$ and applies local 
functions to produce an output that is represented as $A',F',K'$.
Additionally, for any new attribute produced by a UDF (in the output schema $A'$),
its dependencies on the input (in terms of $A,F,K$) and are recorded as a signature 
along with the unique UDF-name.

\onefig{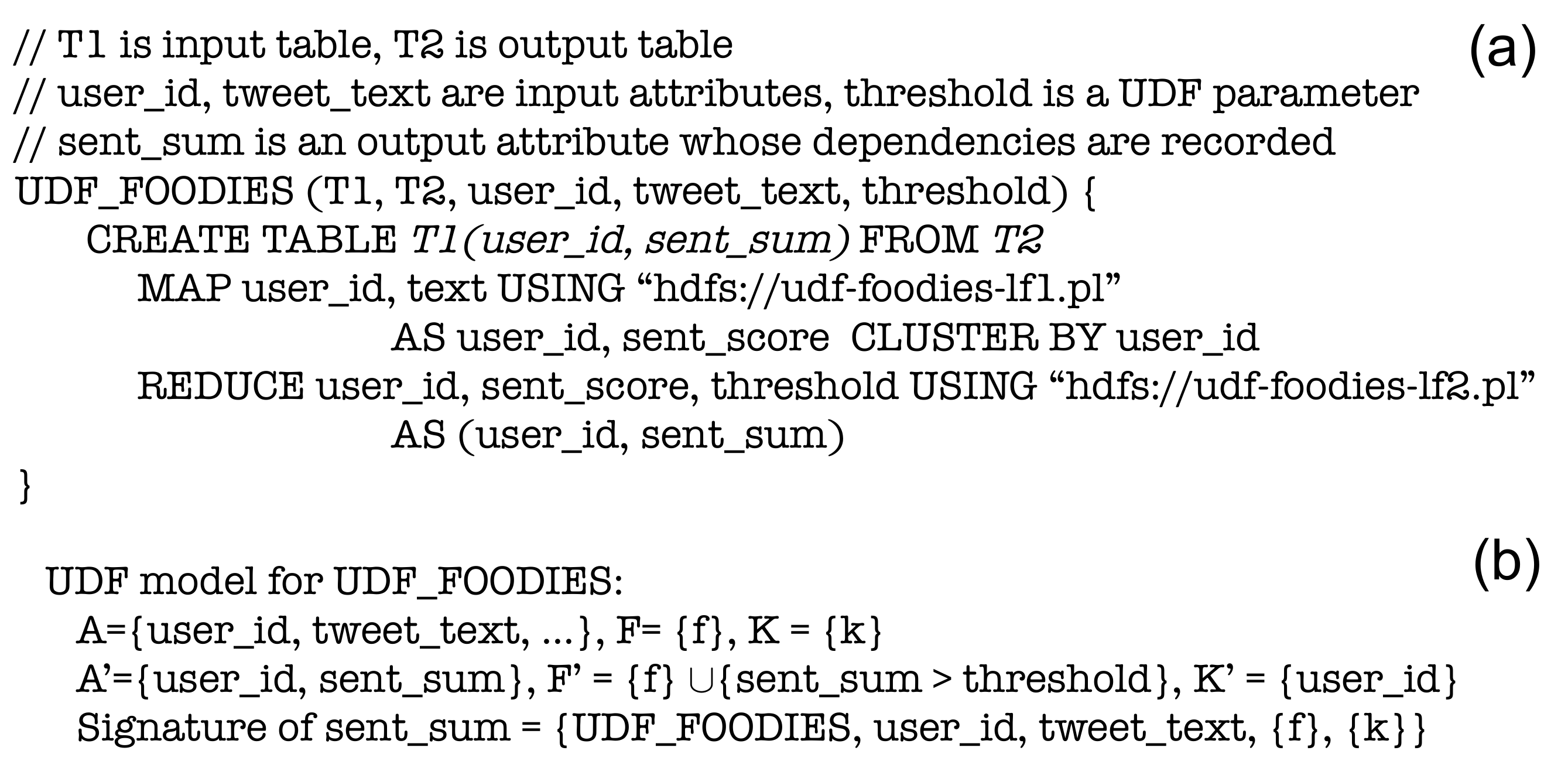}
{UDF\_FOODIES a) implementation composed of two local functions,
b) UDF model showing the end-to-end transformation of input to 
output.}{3.50in}

As an example, consider UDF\_FOODIES that applies a food sentiment
classifier on tweets to identify users that tweet positively about food. 
An abbreviated HiveQL definition of the UDF is given in 
Figure~\ref{fig-composition-udf-local-function-example}(a) that invokes
the following two local functions lf$_1$ and lf$_2$
written in a high-level language (Perl in this example).
lf$_1$: For each ({\tt user\_id}, {\tt tweet\_text}),
apply the food sentiment classifier function that computes a 
sentiment value for each tweet about food.
lf$_2$: For each {\tt user\_id}, compute the sum of the sentiment values to
produce {\tt sent\_sum}, then filter out users with a total score 
greater than a {\tt threshold}.

The two local functions correspond to arbitrary 
user code that perform complex text processing tasks such as 
parsing, word-stemming, entity tagging, and word sentiment scoring. 
Yet, the UDF model succinctly
captures the end-to-end transformation of this complex UDF 
as shown in Figure~\ref{fig-composition-udf-local-function-example}(b).
In the figure, the end-to-end transformation of {\tt UDF\_FOODIES} is
captured by recording the changes made to the input $A$, $F$ and $K$ 
by the UDF functions that produces $A'$, $F'$ and $K'$ using a simple notation.
Furthermore, 
for the new attribute {\tt sent\_sum} in $A'$, its dependencies on the subset
of the inputs are recorded.
We provide a more concrete example of the application of the UDF model in a 
HIVEQL query in Section~\ref{sec-udf-example}.
In this way, the model encodes arbitrary user-code representing a sequence of 
MR jobs, by only capturing its end-to-end transformations.

Our approach represents a gray-box  model for UDFs, giving the system a
limited view of the UDF's functionality yet allowing the system to understand 
the UDF's transformations in a useful way.
In contrast, a white-box approach requires a complete understanding of 
\textit{how} the transformations are performed, imposing significant overhead 
on the system.
%For example, without a white-box approach, UDFs with different names 
%performing the same internal computations on the same input will produce 
%results that do not appear identical to the model.
While with a black-box model, there is very little overhead but no semantic 
understanding of the transformations, limiting the opportunity to reuse any 
previous results.

\subsection{Applying the UDF Model and Annotations}\label{sec-udf-example}

\onefigtop{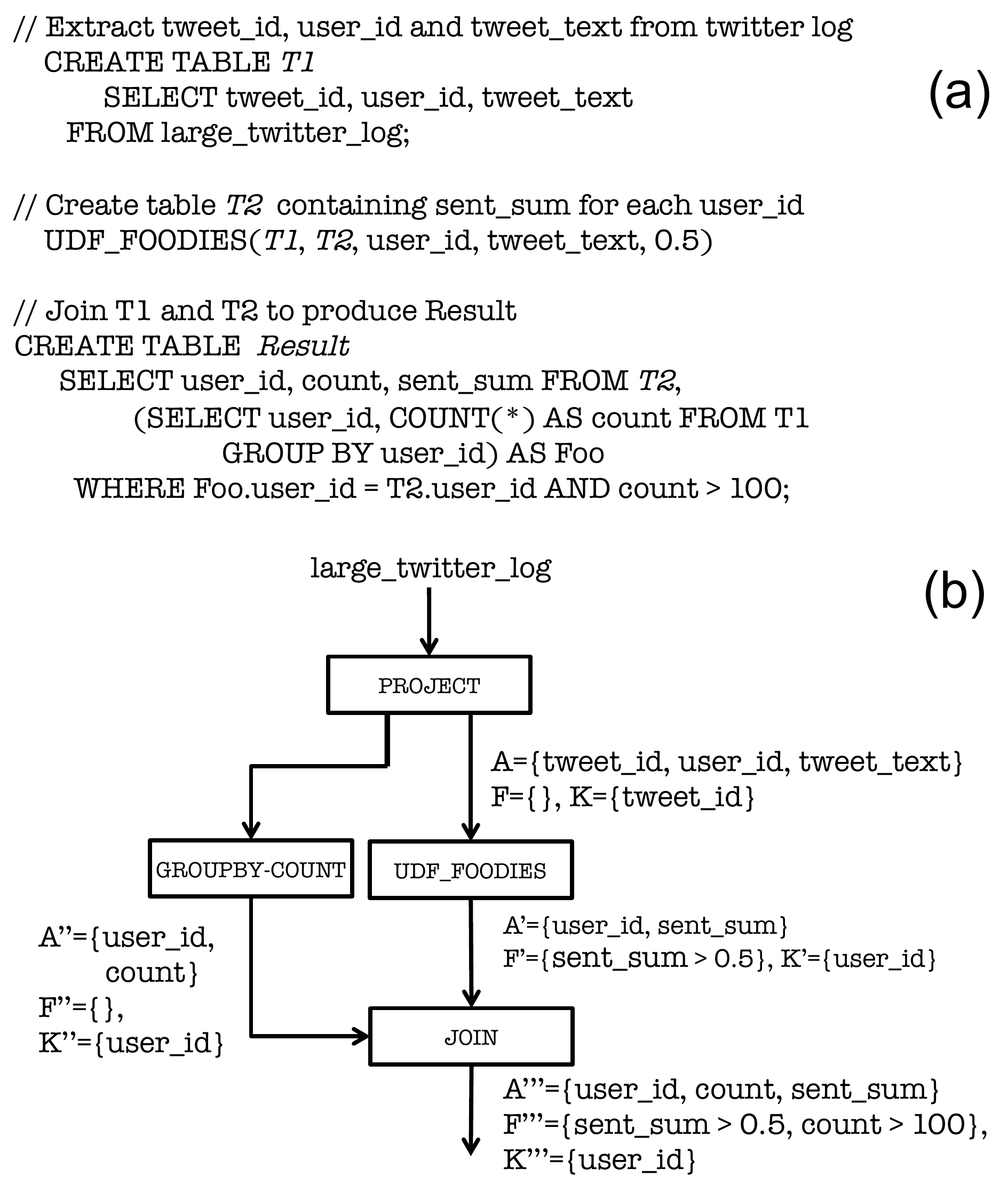}{(a) Example query to obtain prolific 
foodies, and (b) corresponding annotated query plan.
%  , and (c) attribute  dependencies.
}{3.25in}

Having presented our model for UDFs, we now show how to use it
to annotate a query plan that contains both UDFs and relational operators.
In Figure~\ref{fig-udf-example-query}(a), we show a 
query that uses Twitter data to identify prolific 
users who talk positively about food (i.e., ``\textit{foodies}''). 
%This example query makes use of the {\tt UDF\_FOODIES} as well as relational operations.
%and describe how to apply the model to an example UDF. 
The query is expressed in a simplified representation of HiveQL and
applies \texttt{UDF\_FOODIES} from Figure~\ref{fig-composition-udf-local-function-example}(a)
that computes a food sentiment score  ({\tt sent\_sum}) per user based on each user's tweets.

The HiveQL query is converted to an annotated 
plan as shown in Figure~\ref{fig-udf-example-query}(b)
by utilizing the UDF model of  \texttt{UDF\_FOODIES} as given in 
Figure~\ref{fig-composition-udf-local-function-example}(b).
%In this way, the model is not encoding a simple SPJGA query but rather 
%arbitrary user-code representing a sequence of MR jobs.
In addition to modeling UDFs, the three operations 
(denoted as 1, 2, 3 above) can also be used to characterize standard relational 
operators such as select (2), project (1), join (2,3), group-by (3), and 
aggregation (3,1).
Joins in MR can be performed as a grouping of multiple relations on a common key 
(e.g., co-group in Pig) and applying a filter. 
Similarly, aggregations are a re-keying of the input (reflected in $K'$) 
 producing a new output attribute (reflected in $A'$).
%Figure~\ref{fig-udf-example-query} described next.
These $A,F,K$ annotations can be applied to both UDFs and relational 
operations, enabling the system to automatically annotate
every edge in the query plan.

Figure~\ref{fig-udf-example-query}(b) shows the input to the UDF is modeled as 
$\langle A$=\{\texttt{user\_id}, \texttt{tweet\_id}, \texttt{tweet\_text}\},
$F$=$\emptyset$,
$K$=\texttt{tweet\_id}$\rangle$.
The output is 
$\langle A'$=\{\texttt{user\_id}, \texttt{sent\_sum}\}, 
$F'$=\texttt{sent\_sum} > 0.5,
$K'$=\texttt{user\_id}$\rangle$.
\texttt{UDF\_FOODIES} produces the \emph{new} attribute   
\texttt{sent\_sum} whose dependencies are recorded (i.e., signature) as: 
$\langle A$=\{\texttt{user\_id}, \texttt{tweet\_text}\},
$F$=$\emptyset$,
$K$=\texttt{tweet\_id}, udf\_name=\texttt{UDF\_FOODIES}$\rangle$.
%Each signature is recorded by the system as shown in Figure~\ref{fig-udf-example-query}(c).
Lastly, as shown in Figure~\ref{fig-udf-example-query}(b), the output of the 
UDF ($A',F',K'$) forms one input to the subsequent join operator, which 
in turn transforms its inputs to the final result. 

This example shows how a query containing a UDF with arbitrary user code
can be semantically modeled. % and potentially reused.
The $A,F,K$ properties are straightforward and can be provided as annotations 
by the UDF creator with minimal overhead, or alternatively they may 
be automatically deduced via some code analysis method such as~\cite{Hues2012}.
The annotations for each UDF are only provided once, i.e, the first time the 
UDF is added to the system.

While this model may appear limited in its expressiveness, in practice it 
captures a large class of common UDFs.
As an example, we performed an empirical analysis of two real-world UDF libraries,    
Piggybank~\cite{piggybank} and DataFu~\cite{datafu}.  
Our model captures 90\% of the UDFs examined: 
16 out of 16 Piggybank UDFs, and 30 out of 35 DataFu UDFs detailed 
in~\cite{Lefe2013}.
Two classes of UDFs not captured by our model are: 
(a) non-deterministic UDFs such as those that rely on runtime properties 
(e.g., current time, random, and stateful UDFs) and 
(b) UDFs where the output schema itself is dependent upon the input data values 
(e.g., pivot UDFs, contextual UDFs).  

%In this work, the user provides simple annotations only the first time the UDF
% is admitted to the system.  
%Finally, in our model, UDFs are stateless and self-contained like most
%``reasonable'' UDFs in the MR framework, whereas stateful and 
%non-deterministic UDFs are outside the scope of our work.
%%Our model does not capture stateful UDFs with side-effects, as these are not
%typical in MapReduce.  
%However, we do not restrict the use of such UDFs but the system will 
%view them as black-boxes, limiting their opportunities for reuse. 

\section{Using the UDF Model to Perform Rewrites}\label{sec-using-model}

Our goal is to  leverage previously computed results when answering a new 
query.
The UDF model aids us in achieving this goal in three ways:   
First, it provides a way to check for equivalence.
Second, it aids in the costing of UDFs.
Third, it provides a lower-bound on the cost of a potential rewrite.

\subsection{Equivalence Testing}\label{sec-equivalence}

%If we consider the annotated edges of a plan $W$ as views, then 
The system searches for rewrites using existing views and can test for semantic
equivalence in terms of our model using the properties $A$, $F$, and $K$.
We consider a query and a view to be equivalent if they have
identical $A$, $F$ and $K$ properties.
If a query and a view are not equivalent, our system considers applying transformations 
(sometimes referred to as compensations) to make the existing view equivalent
to the query.

Here we develop the mechanics to test if a query $q$ 
(i.e., a target in the annotated plan) can be rewritten using 
an existing view $v$.
Query $q$ can be rewritten using view $v$ if $v$ \textit{contains} $q$.
Checking containment is a known hard problem~\cite{Jayr2006} even for 
conjunctive queries, hence we make a first guess that only serves 
as a quick conservative approximation of containment.
This conservative guess allows us to focus computational efforts toward 
checking containment on the most promising previous 
results and avoid wasting computational effort on less promising ones.

We provide a function $\complete(q,v)$ that performs this heuristic check.
%which we describe next.
$\complete(q,v)$ takes an optimistic approach, representing a \emph{guess} 
that a complete rewrite of $q$ exists using only $v$.
This guess requires the following necessary conditions
as described in~\cite{Hale2001} (SPJ) and~\cite{Gold2001} (SPJGA)
that a view must satisfy to participate in a complete rewrite of $q$.  % ,
%although these conditions are not sufficient to
%confirm the existence of an equivalent rewrite using $v$.

\begin{enumerate}
\renewcommand{\labelenumi}{(\roman{enumi})}
\item $v$ contains all attributes required by $q$; or contains all 
    necessary attributes to produce those attributes in $q$ that are not in $v$
\item $v$ contains weaker selection predicates than $q$
\item $v$ is less aggregated than $q$
\end{enumerate}

The function $\complete(q,v)$ performs these checks and returns true if $v$
satisfies the properties i--iii with respect to $q$.
Note these conditions under-specify the requirements for determining that a 
valid rewrite exists, as they are necessary but not sufficient conditions.
Thus the guess may result in 
a false positive, but will never result in a false negative.
The purpose of $\complete(q,v)$ is to provide a quick way to distinguish 
between views that can possibly produce a rewrite from views that cannot.
Since rewriting is an expensive process, this helps to avoid examining views 
that cannot produce valid rewrites. 

\subsection{Costing a UDF}\label{sec-costing-udfs}

Given that our goal is to find a low cost rewrite for queries containing UDFs,
we require a method of costing a MapReduce UDF.
% However, since UDFs can apply complex computations to data and contain arbitrary 
% user code it is unrealistic to develop a general cost model for  UDFs.
We define the cost of a UDF as the sum of the cost of its local functions.
Estimating the cost of a local function that performs any of the three 
operation types is complicated by two factors: 

\begin{enumerate}
\renewcommand{\labelenumi}{(\alph{enumi})}

\item Each operation type is performed by arbitrary user code,  and 
they could be of varying complexity.
For instance, consider an NLP sentence tagger and a simple word-counter function.
Although both functions
perform the same operation type (discard or add attributes), they can have
significantly different computational costs.

\item There could be multiple operation types performed in the same local 
function, making it unrealistic to develop a cost model for every possible 
local function.
So, we desire a conservative way to estimate
the cost of a local function that applies a sequence of operations
without knowing how these operations interact with each other 
inside the local function.

\end{enumerate}

Developing an accurate cost model is a general problem for any database system.
In our framework, the importance of the cost model is only in guiding the exploration
of the space of rewrites.
For this reason, we appeal to an existing 
cost model from the literature~\cite{Nyki2010}, 
but slightly modify it to be able to cost UDFs.
However, an improved cost model can be plugged in as it becomes available.
Here, we develop a simple cost model that works well in practice.
%Hence a ``data only'' cost model such as that in~\cite{Nyki2010} that does not 
%model computation cost is not sufficient for our purposes.
To this end, we extend the ``data only'' cost model in~\cite{Nyki2010} in a 
limited way so that we are able to produce cost estimates for UDFs. 
Although this results in a rough cost estimate, experimentally we show that 
our cost model is effective in producing low cost rewrites 
(Section~\ref{sec-experimental-evaluation}).

Recall that UDFs are composed of local functions, where each local function 
must be performed by a map task or reduce task.
The cost model in~\cite{Nyki2010} accounts for the ``data'' costs 
(read/write/shuffle), and  we augment it in a limited way to account 
for the ``computational'' cost of local functions.
Since  a UDF can encompass multiple jobs, we express the cost of each job as 
the sum of:  
the cost to read the data and apply a map task ($C_{m}$),
the cost of sorting and copying ($C_{s}$),
the cost to transfer data ($C_{t}$),
the cost to aggregate data and apply a reduce task ($C_{r}$),
and finally the cost to materialize the  output ($C_{w}$).
Using this as a {\em generic} cost model, we first describe 
our approach toward solving (a) by assuming that each
local function only performs one instance of a single operation type.
Then we describe our approach for (b). 
%, and put these together for our solution.

For (a) we model the cost of
the three operation types rather than each local function.
This gives a baseline calibration step for each operation type. %, common for any system.
% However, the direct application of the baseline cost model to any arbitrary local function
% is obviously not reasonable since there can be a high variance in computational costs.
% For instance, producing a new attribute with an NLP sentence tagger function has a far higher cost 
% than producing a new attribute with a simple word-counter function, 
%although both belong to the same operation type (add/remove an attribute).
% although both belong to the same operation type (discard or add attributes).
As noted, there can be a high variance in computational costs of local functions
that perform the same operation type.
To remedy this, the computation cost of every local function 
(assuming for now it only performs a single operation type), 
is initially set to the cost provided by the generic model.
% This serves as the base cost model, 
% then 
For a local function that is computationally expensive,
we scale-up its cost by applying a scalar multiplier to the generic cost of
$C_m$ and $C_r$.
Identifying this scalar value is a one-time process that we obtain by running 
a micro-job on a small sample of its input data, performed only
the first time the UDF is added to the system. 

%The types of operations performed by a local function can be deduced from the 
%way HiveCLI requires UDFs to be given as a sequence of 
%local functions, where each function must specify whether it is executed as 
%a map or reduce task.

For (b), since a local function performs an arbitrary sequence
of operations of any type, it is difficult to estimate its cost.
This would require knowing how the different operations actually interact with one another
as in a white-box approach.
% Without a white-box approach, it is difficult to estimate the computational 
% cost of any local function that performs several operation types.
For this reason we desire a conservative way to estimate the cost of a local function,
which we do by appealing to the following property of any cost model 
performing a \emph{set} $S$ of operations.

\begin{definition}\label{definition-non-subsumable-property}
Non-subsumable cost property:
Let $\cost(S,D)$ be defined as the total cost of performing all operations 
in $S$ on a database instance $D$.
% Let $\cost(x,D) = \infty$ for those operations $x \in S$ that
% cannot be applied on $D$.
The cost of performing $S$ on a database instance $D$ is at least
as much as performing the cheapest operation in $S$ on $D$.
$$\cost(S,D) \geq \min(\cost(x,D), \forall x \in S) $$ 
\end{definition}

% Could use MR-cost model or a RDBMS cost model depending on the underlying store
% SJ: Need to extend it to support \ocost functionality
%Taking the minimum cost of an operation in $S$ provides a lower bound on the
%cost of $S$ given $D$.
%Because the lower bound is derived from the minimum operation cost, it provides a weak lower bound.
%Noting the fact that at least one operation in $S$ must be applied to $D$ allows us to use the cost
%of the least expensive operation in $S$.
The gray-box model of the UDFs only captures enough information about the local
 functions to provide a cost corresponding to the least expensive operation 
performed on the input.
% A tighter bound might be provided if we could use the cost of the most expensive operation in $S$,
% i.e., $\max(\cost(x,D), \forall x \in S)$.
We cannot use the most expensive operation in $S$ 
(i.e., $\max(\cost(x,D), \forall x \in S)$), since this requires 
$\cost(S', D) \leq \cost(S,D)$, where $S' \subseteq S$.
The ``max'' requirement is difficult to meet in practice, 
which we prove using a simple counter-example.
Suppose $S$ contains a filter 
with high selectivity, and a group-by with higher cost than the filter 
when considering these operations independently on database $D$.
Let $S'$ contain only group-by.
Suppose that applying the filter before group-by results in few or no tuples 
streamed to group-by.
Then applying group-by can have nearly zero cost and it is plausible that
$\cost(S', D) > \cost(S,D). \qquad\blacksquare$ 
     
We utilize the non-subsumable cost property in the following way.
A local function that performs multiple operation types $t$ is given 
an initial cost corresponding to the generic cost of applying the cheapest operation 
type in $t$ on its input data.
This initial value can then be scaled-up as described previously in our 
solution for (a). \looseness=-1

\subsection{Lower-bound on Cost of a Potential Rewrite}\label{sec-ocost}

Now that we have a quick way to determine if a view $v$ can potentially 
produce a rewrite for query $q$,  and a method for costing UDFs, 
we would like to compute a quick lower bound on the cost of any
\textit{potential} rewrite -- without having to actually find a valid rewrite, 
which is computationally hard.
To do this, we will utilize our UDF model and the non-subsumable cost property 
when computing the lower-bound.  
The ability to quickly compute a lower-bound is a key feature of our approach.
% that is critically important to our practical solution.

%Suppose that $v$ is guessed to be complete with respect to $q$, and $r$ 
%denotes a valid rewrite of $q$ using $v$.
%Given $v$ as input, the transformations performed by $r$ must produce 
%an output that is equivalent to $q$.
%Notice that using the UDF model, the representation of $q$ in terms of $A,F,G$
%can be compared with that of $v$ in terms of $A,F,G$.
%Then the transformation produced by $r$ is given by the set difference between the attributes, filters and group-bys representation 
%of $q$ and $v$ and is referred to as the {\em fix}.

\onefig{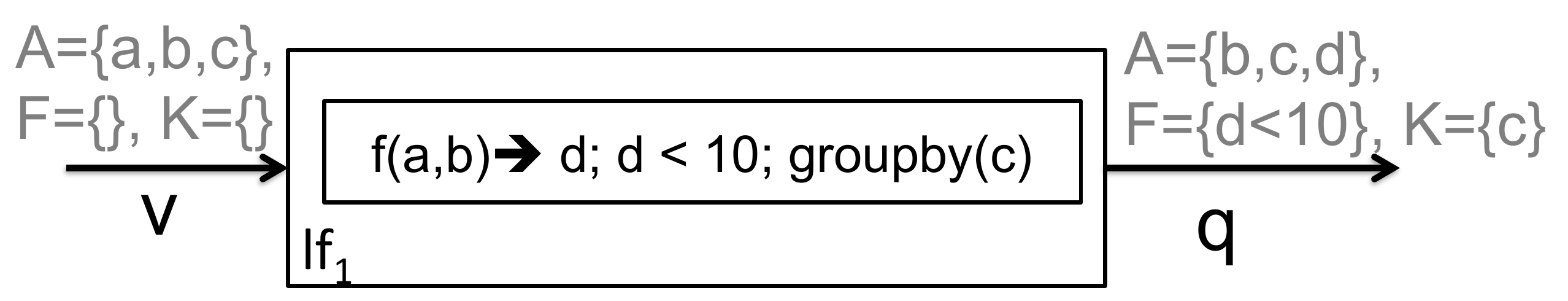}{Synthesized UDF to perform the fix between a view $v$ and a query $q$.}{3in}

As an example of computing the lower bound, we show a view $v$ and a query 
$q$ in Figure~\ref{fig-example-fix} annotated using the model.
Suppose that $v$ is given by attributes $\{a,b,c\}$ with no applied filters 
or grouping keys. 
Now, suppose that $q$ is given by $\{b,c,d\}$, has a filter $d < 10$, and has 
key $c$, where attribute $d$ is computed using $a$ and $b$, which happen to be 
present in $v$.
It is clear that $v$ is guessed to be complete with respect to $q$ because
$v$ has the required attributes to produce $q$ and $v$ has weaker filters 
and grouping keys (i.e., is less aggregated) than $q$.
Note that the guess implies it may not be complete since it may be possible
that the application of the grouping on $c$ may remove $a$ and $b$, rendering 
 the creation of $d$ not possible.
However, since it passes the $\complete(q,v)$ test, we can then compute what 
we term the \textit{fix} for $v$ with respect to $q$.
Using the UDF model, the representation of $q$ in terms of $A,F,K$
can be compared with that of $v$ in terms of $A,F,K$.
To compute the fix, we only take the set difference between the attributes,
filters, and group-bys ($A,F,K$), which is straightforward and simple to 
compute.
In Figure~\ref{fig-example-fix}, the fix for $v$ with respect to $q$ is given
by: a new attribute $d$; a filter $d < 10$; and group-by on $c$.

To produce a valid rewrite we need to find a sequence of local functions that ``perform'' the fix; these are the operations that when applied to $v$ will 
produce $q$. 
%Note this is the same as find a valid rewrite, which is a known hard problem,
%Finding  this sequence of local functions is the same problem as finding a 
%valid rewite, which is a known hard problem.
As this a known hard problem, we \textit{synthesize} a hypothetical UDF 
comprised of a single local function that applies all operations in the fix.
The cost of this synthesized UDF, which serves as an initial stand-in for a 
potential rewrite should one exist, is obtained using our UDF cost model.
%If the fix is non-empty, then at least one local function must be applied 
%to produce a valid rewrite.
%This local function must apply all operations in the fix.
%We can obtain the cost of this local function using our UDF cost model.
This cost corresponds to the lower-bound of any valid rewrite $r$ --- by the 
non-subsumable cost property, the computational cost of this single
local function is the cost of the cheapest operation in the fix.
The benefit of the lower-bound is that it lets us cost views by their potential 
ability to produce a low-cost rewrite, without having to expend the computational
effort to actually find one.
Later we show how this allows us to consider views that are ``more promising'' 
to produce a low-cost rewrite before the ``less promising'' views are
considered.

We define an optimistic cost function $\ocost(q,v)$ that computes this 
lower-bound on any rewrite $r$ of query $q$ using view $v$ only if 
$\complete(q,v)$ is true.
Otherwise $v$ is given \ocost{} of $\infty$, since in this case it cannot 
produce a complete rewrite, and hence the \cost{} is also $\infty$.
%a rewrite cannot be found thus the \cost{} is also $\infty$.
The properties of $\ocost(q,v)$ are that it is very quick to compute and
%(i.e., does not require actually finding a rewrite and then obtaining its cost), 
$$\ocost(q,v) \leq \cost(r).$$
When searching for the optimal rewrite $r^*$ of $W$,  
we use $\ocost{}$  to enumerate the space of the candidate 
views based on their cost potential, as we describe in the next section.
This is inspired by nearest neighbor finding problems
in metric spaces where computing distances between objects can be 
computationally expensive, thus preferring an alternate distance function 
(e.g., $\ocost{}$) that is easy to compute with the desirable property that 
it is always less than or equal to the actual distance.

\section{Problem Overview for Rewriting Queries Containing UDFs}
\label{sec-problem-definition}

%  Why is this problem hard?
%  Determining if a rewrite exists
%  Determining a candidate rewrite
%  Containment 

Our UDF model enables reuse of views to improve query performance 
even when queries contain complex functions.
However, reusing an existing view when rewriting a query with any arbitrary UDF
requires the rewrite process to consider all UDFs in the system.
%Hence any UDF admitted to the system becomes part of the rewrite language.
Given that the system will likely have many users who write a lot of queries
and UDFs, including all these UDFs 
in the rewrite process makes searching for the optimal rewrite impractical for
any realistic workload and number of views.  
This is because the search space for finding a rewrite is exponential in both 
1) the number of views in $V$ and 2) the operations considered by 
the rewrite process, which may include multiple applications of the same
operator.
The problem is known to be hard even when both the queries and the 
views are expressed in a language that only includes conjunctive 
 queries~\cite{Hale2001,Abit1998,Levy1995}.

%%%%%%%%%%%%%%%%%%%%%%%%%%%%%%%%%%%%%%%%%%%%%
%% IMPORTANT Comment from Carey.
%% KEEP  %%%%%%%%%%%%%%%%%%%%%%%%%%%%%%%%%%%%
\begin{comment}
Since a single query typically only have  a few udfs, why not restrict
the rewrite language to include SPJGA+the-UDFs-In-The-Query.  
This makes the rewrite language the same as the query language,
hence all you need to do is solve 1 rewrite search problem corresponding
to $W_n$.
This can be done by unfolding all of the UDF variable signatures to 
obtain their operators and UDFs have been applied to the input.
This is a reasonable solution, however, the 1 rewrite search problem
could have a very large space.  
Suppose a query has 15 targets, then the single problem would have
15! possible ordering of the operators you unfolded.
But in our case, it will be 15 rewrite search problems each with a small
size (e.g. SPJGA+UDF_1, 6 operations) would be 6!, resulting complexity
is 15*6!, and this corresponds to a naive rewrite solution, whereas our
BFR shows you can often much better than this. Hence:
15! >> 15*6!
\end{comment}
%% KEEP  %%%%%%%%%%%%%%%%%%%%%%%%%%%%%%%%%%%%
 
In our system, both the queries and the views can contain any arbitrary UDF.
For  practical reasons it is necessary that only a small subset of all UDFs be
  considered by the rewrite process.  
Our rewriter considers relational operators --- select, project, join, 
group-by, aggregations (SPJGA), and a few of the most frequently used 
UDFs, which increases the possibility of reusing previous results. 
%Our system provides an easy way of adding new UDFs, but these are not 
%automatically known to the rewriter; 
Selecting the right subset of UDFs to include in the rewrite process 
is an interesting open problem that must consider the tradeoff between the
 added expressiveness of the rewrite process versus the additional exponential
cost incurred to search for rewrites.

%% Note if you do not include all UDFs as noted above, then you must  
% do n search problems to find r*
Finding an optimal rewrite of only $W_n$ does not suffice, as  
we will illustrate below in Example~\ref{ex-greedy-plumbing}.
This is because limiting the rewrite process to include only a subset of UDFs 
means that finding the optimal rewrite for $W$ requires that we solve $n$ 
rewrite search problems, i.e., one for each of the $n$ targets 
(Section~\ref{sec-preliminaries-definitions}) in $W$.
%This is because finding an optimal rewrite of only $W_n$ does not suffice as 
%we illustrate next in Example~\ref{ex-greedy-plumbing}.
Since the rewriter does not consider all UDFs, even if one cannot find a 
rewrite for $W_n$, one \textit{may} be able to find a rewrite at a 
different target in $W$.
Furthermore, even if a rewrite is found for $W_n$, there may be a 
\textit{cheaper} rewrite of $W$ using a rewrite found for a different 
target $W_i$.
For example, a rewrite $r_i$ found for $W_i$  can be expressed as a rewrite
for $W_n$ by combining $r_i$ with the remaining nodes in $W$
indicated by $\node{i+1} \cdots \node{n}$.
Thus the search process for the optimal rewrite must happen at all $n$ targets 
in $W$. 
%, which complicates a solution since solving even a single instance of 
%the problem is hard.

A naive solution is to search for rewrites at all $n$ targets of $W$
completely independently.
This approach finds the best rewrite for each target, if one exists, and
chooses a subset of these to obtain the optimal rewrite $r^*$.
One drawback of this approach is that there is no way of terminating early the
 search at a single target.
Another drawback is that even with an early termination property, the
algorithm may search for a long time at a target (e.g., $W_i$) only to find an 
expensive rewrite, when it could have found a better (lower-cost) rewrite at 
 an upstream target (e.g., $W_{i-1}$) more quickly if it had known where to 
 look first. 
This is illustrated in Example~\ref{ex-greedy-plumbing}.

\begin{example}\label{ex-greedy-plumbing}
$W$ contains 3 MR job nodes, $n_1$, $n_2$, $n_3$, each with their individual 
node cost as indicated, where the total cost of $W$ is 13 (6+5+2).
Alongside each node is the space of views ($V$) to consider for rewriting.
\begin{wrapfigure}{r}{1.15in}
\centering
\vspace{-0.1in}
\hspace*{-0.25in}\includegraphics[width=1.5in]{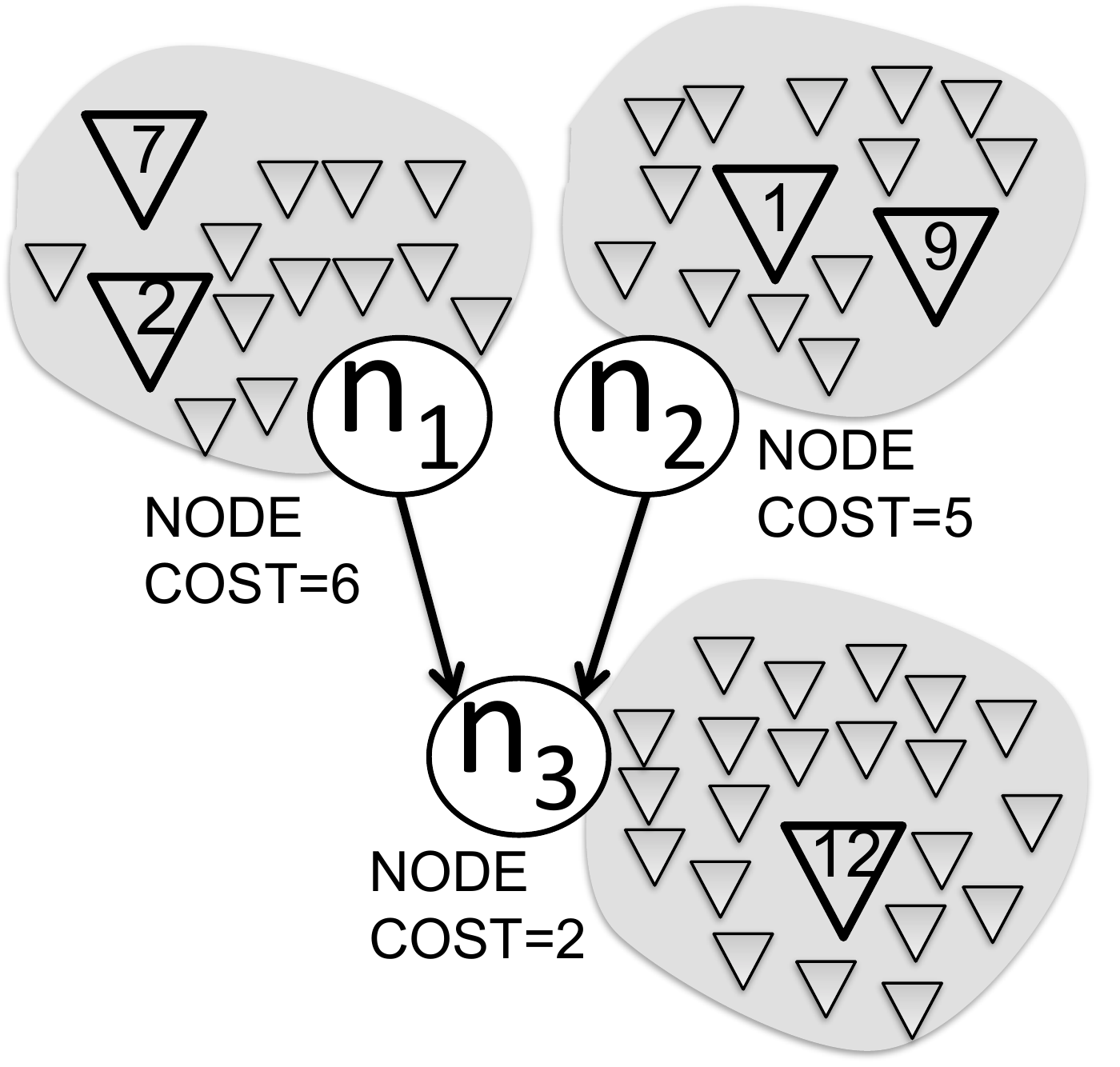}\hspace*{2cm}
\vspace{-0.1in}
\end{wrapfigure}
Candidate views that fail to yield a rewrite are indicated by the empty triangles,  
and those that result in a rewrite are indicated by cost of the rewrite found.
A naive algorithm would first examine \emph{exhaustively} the views at $n_3$,
 finally identifying the rewrite of $W$ with a cost of $12$.
However, as noted, to find the optimal rewrite it cannot stop at this point,
and must continue searching for rewrites at $n_1$ and $n_2$.
The algorithm would then find a rewrite at $n_1$ of cost 2, and at $n_2$ of cost 1.
It then combines these with the node $n_3$ (node cost 2), 
resulting in a rewrite of $W$ with a total cost of 5 (2+1+2).
This is much less than the earlier rewrite found at $n_3$ with a cost of 12.
This example shows the algorithm cannot stop even when it finds a rewrite for 
$n_3$.
Also, had the algorithm known about the low-cost rewrites at $n_1$
and $n_2$, it need not have exhaustively searched the space at $n_3$.
\end{example}

\subsection{Overview of our Approach}\label{sec-solution-outline}
We can improve the search overhead for the optimal rewrite 
by making use of the lower bound function $\ocost{}$ introduced in 
Section~\ref{sec-ocost}.
During the search for rewrites at each of the $n$ targets, 
the lower bound 
can be used to help terminate the search earlier in two ways.
First, for any rewrite $r$ found at a given target, if 
the lower-bound on the cost of any possible  rewrites remaining in the 
unexplored space is greater than the cost of $r$, there is no need 
to continue searching the remaining space. 
This enables us to terminate the search early at a single target.
Second, we can use $r$ and the lower bound on the remaining unexplored 
space at one target to inform the search at a different target.
For instance, in Example~\ref{ex-greedy-plumbing}, after finding 
the best rewrites for $n_1$ and $n_2$ of cost 2 and 1 respectively, 
we can stop searching at $n_3$ when the lower-bound on the cost of the
unexplored space at $n_3$ is greater than 5, since we already have 
found a rewrite of $W$ with total cost 5.

%This approach is clearly inefficient.
%Suppose there is an oracle that could reveal the best rewrite for each target, 
%as well as its cost.
%An algorithm could make use of the oracle in two ways.
%First it could determine the correct order of visiting targets.
%For instance, rewrites for $n_1$ and $n_2$ should be considered before $n_3$
%as they have the potential to create a low cost rewrite for $n_3$, as shown in 
%the example.
%Second it could prune some of the targets from consideration.
%For instance, there is no need to search for rewrites at $n_3$ that have a cost
%greater than 4.
%Since the oracle knows the best rewrite of $n_3$ has a cost of 12,
%the algorithm can prune $n_3$ from the search space.

We propose a work-efficient query rewriting algorithm that uses $\ocost{}$ to 
order the search space at each target.
Since $\ocost{}$ is easy to compute, it enables us to quickly order the 
candidate views at each target by the lower-bound
on their ability to produce a rewrite (if one exists).
This allows our algorithm to step through the space at each target in an 
incremental fashion.
%Our approach is inspired by nearest neighbor finding problems
%in metric spaces where computing distances between objects can be 
%computationally expensive, thus preferring an alternate distance function 
%(e.g., $\ocost{}$) that is easy to compute with the desirable property that 
%it is always less than or equal to the actual distance.

Using the $\ocost{}$ function to order the space, our rewrite algorithm finds the 
optimal rewrite $r^*$ of $W$ by breaking the problem into two components:
\begin{enumerate}
\item $\WF{}$ (Section~\ref{sec-best-first-workflow-rewrite})
% that tries to combine the rewrites found for all the $n$ targets into
performs an efficient search of rewrites for all targets in $W$ and outputs a
globally optimal rewrite for $W_n$.
\item $\VF{}$ (Section~\ref{sec-viewfinder}) enumerates candidate views 
for a single target based on their potential to produce a low-cost rewrite of 
the target, and is utilized by $\WF{}$.
\end{enumerate}

\section{Best-First Rewrite}\label{sec-best-first-workflow-rewrite}

The $\WF{}$ algorithm produces a rewrite 
$r^∗$ of $W$ that can be composed of rewrites found at multiple targets in $W$. 
The computed rewrite $r^*$ has provably the \emph{minimum cost} among all possible rewrites in the same class. 
Moreover, the algorithm is \emph{work-efficient}: even though $\cost(r^*)$ is not known a-priori, it will never examine any candidate view with $\ocost{}$
higher than the optimal cost $\cost(r^*)$. 
Intuitively, the algorithm explores only the part of the search space that 
is needed to provably find the optimal rewrite.
We prove that $\WF{}$ finds $r^*$ while being work-efficient in 
Section~\ref{sec-wf-proof}.

\begin{comment}
The $\WF{}$ algorithm produces an {\em optimal} rewrite $r^*$ of 
a plan $W$ in a  {\em work efficient} manner.
The rewrite $r^*$ is {\em optimal} as there cannot be a cheaper
rewrite for $W$ than $r^*$, where $r^*$ can be composed of rewrites found at multiple targets in $W$.
% The proposed algorithm is also {\em work efficient} in the sense
% that it will never explore more of the solution space than the minimum
% necessary to find $r^*$.
We formally define the work efficiency property as follows:
$\WF{}$ will not examine any candidate view with $\ocost{}$ greater than $\cost(r^*)$.
We prove both these properties of our algorithm in Section~\ref{sec-wf-proof}.
\end{comment}

The algorithm begins with $W$ itself considered as the best rewrite 
 (i.e., lowest-cost) for the plan. 
It then spawns $n$ concurrent search problems at each of the targets in 
$W$ and works in iterations to find a better rewrite. 
In each iteration, the algorithm chooses one target $W_i$
and examines a candidate view at $W_i$.
The algorithm makes use of the result of this step to aid in pruning the
search space of other targets in $W$. 
To be work efficient,
the algorithm must choose wisely the next candidate view to examine.
As we will show below, the $\ocost$ functionality plays an essential role
in choosing the next target to refine.

%The key feature of our method is the $\ocost$ functionality that allows the $\WF{}$
%to incrementally search the space of rewrites of $W$, such that $\WF$
%finds the optimal rewrite of $W$ in a work efficient manner.

The $\WF$ uses an instance of the $\VF$ at each target to search
the space of rewrites. 
We will describe the details of $\VF$ in Section~\ref{sec-viewfinder}. 
In this section, $\VF$ is used as a black box that provides the following 
functions for each target $W_i$: 
%The $\WF$ treats $\VF$ as a black-box until Section~\ref{sec-viewfinder} 
% and exploits its $\ocost$ functionality
%to incrementally search the space of rewrites for $W$.
%We assume that the $\VF{}$ at each target implements three functions: 
(1) {\sc Init} creates the search space of candidate views ordered by their $\ocost{}$, 
(2) {\sc Peek} provides the $\ocost{}$ of the \textit{next} candidate view, and
(3) {\sc Refine} searches for a rewrite of the target using the next candidate view,
which involves trying to apply the fix.
An important property of {\sc Refine} is the following: 
there are no remaining rewrites to be found for the corresponding
target that have a cost less than the value of {\sc Peek}.
Next in Section~\ref{sec-bfr-algorithm} we describe the $\WF{}$ algorithm in more 
detail and in Section~\ref{sec-bfr-example} we give a small running example
of the algorithm.

\subsection{The $\WF{}$ Algorithm}\label{sec-bfr-algorithm}

\begin{algorithm}[!h]
\caption{Optimal rewrite of $W$ using $\VF$\label{alg-main-setup}}
{\small
\begin{algorithmic}[1]
\Function{$\WF$}{$W$, $V$}
\ForAll{$W_i \in W$}\Comment{Init Step per target}\label{line-wf-init}
  \State $\VF$.{\sc Init}($W_i$, $V$)
  \State $\minPlumb_i$ \qlet $W_i$ \Comment{original plan to produce $W_i$}
  \State $\minCost_i$ \qlet $\cost(W_i)$\Comment{plan cost}
\EndFor \label{line-wf-init-end} \vspace{.2in}
% \State $W_i \qlet W_n$
% \While{$W_i$ $\not= -1$}\label{line-rewrite-wf-start-while-loop}
%   \State $W_i, d_i \qlet \FindNextMinTarget(W_n)$\label{line-wf-find-next-min}
%   \State $\RefineTarget(W_i)$ {\bf if} $W_i \not= -1$
% \EndWhile\label{line-rewrite-wf-end-while-loop}
\Repeat\label{line-rewrite-wf-start-while-loop}
  \State $(W_i, d) \qlet \FindNextMinTarget(W_n)$\label{line-wf-find-next-min}
  \State $\RefineTarget(W_i)$ {\bf if} $W_i \not= \mbox{NULL}$\label{line-wf-refine}
\Until{$W_i$ $= \mbox{NULL}$}\label{line-rewrite-wf-end-while-loop}\Comment{i.e., $d>\minCost_n$}
\State Rewrite $W$ using $\minPlumb_n$
\EndFunction
\end{algorithmic}}
\end{algorithm}

Algorithm~\ref{alg-main-setup} presents the main \WF{} function.
\WF{} first initializes a $\VF$ at each target $W_i \in W$ 
(lines~\ref{line-wf-init}--\ref{line-wf-init-end}),
and initializes $\minPlumb_i$ and $\minCost_i$ as the original plan
and plan cost, respectively.
Then it repeats the following procedure 
(lines~\ref{line-rewrite-wf-start-while-loop}--\ref{line-rewrite-wf-end-while-loop}):
Choose the next best target, $W_i$, to refine with $\FindNextMinTarget$ (line~\ref{line-wf-find-next-min})
given in Algorithm~\ref{alg-find-next-min}; 
then ask \VF{} to refine the next candidate view, with $\RefineTarget$ (line~\ref{line-wf-refine}) 
given in Algorithm~\ref{alg-refine-dag}.

The output $(W_i,d)$ of $\FindNextMinTarget$ means that there is an unexamined 
view at target $W_i$ that can potentially generate a rewrite with a lower-bound
cost of $d$.
% lower-bounded by $d$.
Furthermore, as we will see shortly, $\FindNextMinTarget$ examines views in 
increasing $\ocost{}$ order at each target which guarantees that the return
 value $d$ can never decrease. 
These properties have two implications.  
First, $\WF{}$ can terminate early --- there is no need to continue searching 
if the best rewrite found so far is less than or equal to $d$. 
Second, $\WF{}$ can continue the search at the target with the most 
promising potential rewrite.
The main loop continues 
(lines~\ref{line-rewrite-wf-start-while-loop}--\ref{line-rewrite-wf-end-while-loop}) 
% and updates $\minPlumb_i$
until there is no target that can possibly improve $\minPlumb_n$, at which 
point $r^*$ has been identified.

\begin{comment}
The return value $W_i$ of $\FindNextMinTarget$  corresponds to the next target 
to continue searching, while $d$ represents the minimum \ocost{} of a rewrite 
for $W_n$  involving a candidate view at $W_i$ that has not yet been examined. 
% so far.
As we will see shortly, $\FindNextMinTarget()$ examines views in increasing 
$\ocost{}$
order at each target and so can guarantee that the return value $d$ can never 
decrease. 
This property ensures that $\WF{}$ has examined all possible rewrites for $W$ 
with actual cost less than $d$.
This allows for early termination --- there is no need to continue searching 
if the best rewrite found so far is less than or equal to $d$. 
The main loop iterates % and updates $\minPlumb_i$
until there is no target that can possibly improve $\minPlumb_n$, 
at which point $r^*$ has been identified.
%At this stage of execution \WF{} has  examined all possible rewrites whose 
% actual cost is less than $d$.
\end{comment}

%Next \WF{} invokes Algorithm~\ref{alg-refine-dag} which calls the \Refine{} 
% function of the \VF{} at target $W_i$.
%This asks the \VF{} to examine only the next candidate view for that target,
% as determined by its \ocost{},
%and this forms the basis of the incremental nature of \WF{}.
%While the \VF{} performs an incremental search at each target, the \WF{} 
% performs an incremental search in the space of all possible rewrites for $W$.
%In a sense, $d_i$ never decreases and \WF{} stops when $d_i$ equals 
%$\cost(r^*)$ as we show later.

\begin{algorithm}[!h]
\caption{Find next min target to refine\label{alg-find-next-min}}
{\small
\begin{algorithmic}[1]
\Function{$\FindNextMinTarget$}{$W_i$}
\State  $d' \qlet 0$; $W_{MIN} \qlet \mbox{NULL}$; $d_{MIN} \qlet \infty$
\ForAll{incoming vertex $\node{j}$ of $\node{i}$}\label{line-refine-ancestor-start}
  \State $(W_k, d)$ \qlet $\FindNextMinTarget(W_j)$
  \State $d' \qlet d' + d$
  \If{$d_{MIN} > d$ and $W_k \not= \mbox{NULL}$}\label{line-start-w-min}
    \State $W_{MIN} \qlet W_k$
    \State $d_{MIN} \qlet d$
  \EndIf\label{line-end-w-min}
\EndFor
\State $d'$ \qlet $d' + \cost(\node{i})$\label{line-refine-ancestor-end}
\State $d_i \qlet \VF.\mbox{\sc Peek}()$\label{line-refine-here}
\If{$\min(d',d_i) \geq \minCost_i$} \label{line-nothing-better}
  \State \Return $(\mbox{NULL}, \minCost_i)$
\ElsIf{$d' < d_i$}\label{line-d-i-loses}
  \State \Return $(W_{MIN}, d')$
\Else
  \State \Return $(W_i, d_i)$ \label{line-d-i-wins}
\EndIf
\EndFunction
\end{algorithmic}}
\end{algorithm}

Algorithm~\ref{alg-find-next-min} describes $\FindNextMinTarget{}$ which 
identifies the next best target $W_i$ to be refined in $W$, as well as the 
minimum cost (\ocost) of a potential rewrite for $W_i$.
There can be three outcomes of a search at a target $W_i$.
Case 1: $W_i$ and all its ancestors cannot provide a better rewrite.
Case 2: An ancestor target of $W_i$ can provide a better rewrite.
Case 3: $W_i$ can provide a better rewrite.
By recursively making the above determination at each target $W_i$ in $W$,
the algorithm identifies the best target to refine next.

For a target $W_i$, the cost $d'$ of the cheapest potential rewrite that can be produced by the
ancestors of $\node{i}$ is obtained by summing the \VF.\Peek{} values at $\node{i}$'s ancestors nodes and 
the cost of $\node{i}$ (lines~\ref{line-refine-ancestor-start}--\ref{line-refine-ancestor-end}).
Note that we also record the 
target $W_{MIN}$ representing the ancestor target with the minimum \ocost{} 
candidate view (lines~\ref{line-start-w-min}--\ref{line-end-w-min}).
Next, we assign $d_i$ to the next candidate view at $W_i$
using \VF.\Peek{} (line~\ref{line-refine-here}).
% (1) There is no rewrite better than $\minPlumb_i$,
% (2) A rewrite $r'$ obtained by combining rewrites found at the 
% ancestors of $W_i$ has the best \ocost{} $d'$,
% and (3) A rewrite $r_i$ of $W_i$ using a candidate view 
% at $W_i$ has the best \ocost{} $d_i$.
%

Now the algorithm deals with the three cases outlined above.
If both $d'$ and $d_i$ are greater than or equal to $\minCost_i$ (case 1), 
there is no need to search any further at $W_i$ (line~\ref{line-nothing-better}).
% If both $d'$ and $d_i$ are greater than or equal to $\minCost_i$ this means that $W_i$ and all of its
% ancestors cannot provide a lower cost rewrite than the best rewrite already found for $W_i$
% and there is no need to further refine $W_i$ or any of its ancestors
% for the remainder of the algorithm (i.e., case (1)).
If $d'$ is less than $d_i$ (line~\ref{line-d-i-loses}), then
$W_{MIN}$ is the next target to refine (case 2).
Else (line~\ref{line-d-i-wins}), $W_i$ is the next target to refine (case 3).

%If $d'$ is less than $d_i$, let $W_{MIN}$ represent the ancestor target with the minimum cost candidate view in $r'$,
% and hence must be the next target to refine.
%If $d_i$ is less than or equal to $d'$, then $W_i$ must be the next target to refine.

\begin{algorithm}[!t]
\caption{Queries \VF{} in best-first manner\label{alg-refine-dag}}
{\small
\begin{algorithmic}[1]
\Function{$\RefineTarget$}{$W_i$}
\State $r_i$ \qlet \VF.{\sc Refine}($W_i$)\label{line-obtain-rewrite}
 \If{$r_i \not= \mbox{NULL}$ and $\cost(r_i) < \minCost_i$}\label{line-compare-rewrite}
     \State $\minPlumb_i$ \qlet $r_i$
     \State $\minCost_i$ \qlet $\cost(r_i)$
      \ForAll{edge ($\node{i}$, $\node{k}$)}% \Comment{towards $W_n$}
       \State $\PropagateBestRewrites$($\node{k}$)\label{line-view-finder-propagate-best-plans}
     \EndFor
 \EndIf\label{line-compare-rewrite-end}
\EndFunction
\end{algorithmic}
\Line
\begin{algorithmic}[1]
\Function{$\PropagateBestRewrites$}{$\node{i}$}
\State $r_i$ \qlet plan initialized to $\node{i}$\label{line-rewrite-r-i-start}
\ForAll{edge ($\node{j}$, $\node{i}$)}
  \State Add $\minPlumb_j$ to $r_i$
\EndFor\label{line-rewrite-r-i-end}
\If{$\cost(r_i)$ < $\minCost_i$}\label{line-rewrite-r-i-wins}
   \State $\minCost_i$ \qlet $\cost(r_i)$
   \State $\minPlumb_i$ \qlet $r_i$
   \ForAll{edge ($\node{i}$, $\node{k}$)}
     \State $\PropagateBestRewrites$($\node{k}$)
   \EndFor
\EndIf\label{line-rewrite-r-i-wins-end}
\EndFunction
\end{algorithmic}
}
\end{algorithm}

Algorithm~\ref{alg-refine-dag} describes the process of refining a target $W_i$.
Refinement is a two-step process.
In the first step it obtains a rewrite $r_i$ of $W_i$ from \VF{} if one exists (line~\ref{line-obtain-rewrite}).
The cost of the rewrite $r_i$ obtained by $\RefineTarget$ is compared against the best
rewrite found so far at $W_i$.
If $r_i$ is found to be cheaper, the algorithm suitably updates $\minPlumb_i$ and $\minCost_i$ 
(lines~\ref{line-compare-rewrite}--\ref{line-compare-rewrite-end}).
In the second step (line~\ref{line-view-finder-propagate-best-plans}),
the algorithm tries to compose a new rewrite of $W_n$ using $r_i$, through  
the recursive function given by $\PropagateBestRewrites$ in Algorithm~\ref{alg-refine-dag}.
After this two-step refinement process, $\minPlumb_n$ contains the best rewrite of $W$ found so far.  

The recursion procedure given by $\PropagateBestRewrites$ pushes downward the 
new $\minPlumb_i$ along the outgoing nodes and towards $\node{n}$. 
At each step it composes a rewrite $r_i$ using the immediate ancestor nodes of 
$\node{i}$ (lines~\ref{line-rewrite-r-i-start}--\ref{line-rewrite-r-i-end}).
It compares $r_i$ with $\minPlumb_i$ and updates $\minPlumb_i$ if $r_i$ is 
found to be cheaper 
(lines~\ref{line-rewrite-r-i-wins}--\ref{line-rewrite-r-i-wins-end}).
%
%The recursion pushes the rewrite $r$ downward toward $W_n$, and if it succeeds $r$ becomes part
%of the best rewrite of $W$ found so far.  

\subsection{$\WF$ Algorithm Example}\label{sec-bfr-example}

%Example~\ref{ex-algo-description} shows how \WF{} attempts to rewrite a sample plan.

%%%%%%%%%%%%%%%%%%%%%%%%%%%%%%% Example %%%%%%%%%%%%%%%%%%%%%%%%%%%%%%%%%%%%%%%
%%%%%%%%%%%%%%%%
\onefig{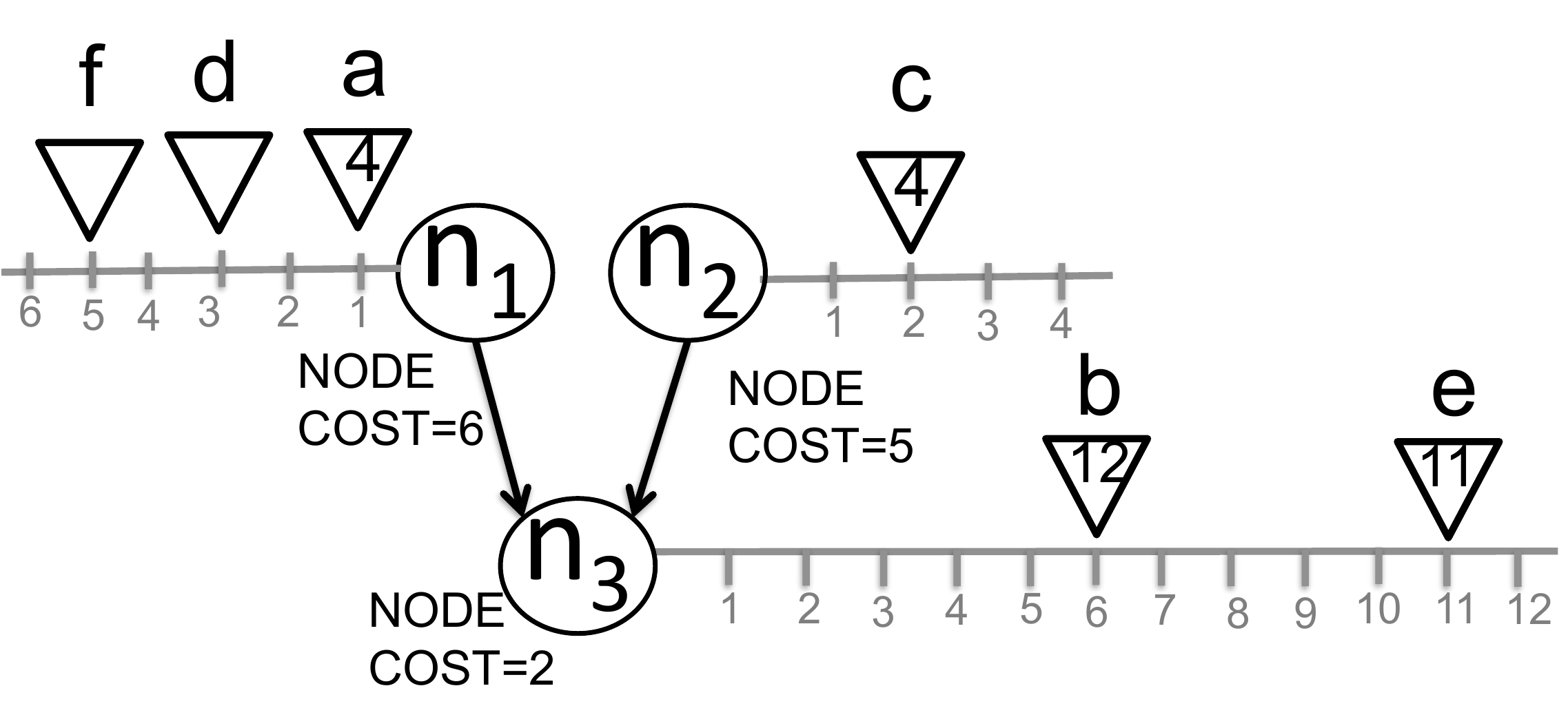}
{Three targets $n_1, n_2$, and $n_3$ with  
candidate views at each target ordered by their $\ocost{}$ $a$--$f$.}
{2.25in}

\begin{example}\label{ex-algo-description}
Figure~\ref{fig-optcost-algo-illustration} shows a plan containing
three nodes $n_1, n_2$ and $n_3$, with node costs 6, 5, and 2 respectively, 
and $n_3$ represents $W_n$.
Therefore, $\minPlumb{}_3$ (i.e.,$\minPlumb{}_n$) begins with \cost{}=13 (6+5+2).
At each target, views ($a$--$f$) are arranged by their 
$\ocost{}$ from their respective target nodes.
For example, $b$ is placed at $\ocost{}$ of 6 for node $n_3$.
An empty triangle in the figure indicates it has not yet been considered by the algorithm; 
initially, all triangles begin empty.
During the course of the example, if it is considered and a rewrite is found, then the actual cost of the 
rewrite is indicated inside.

In the first step of the rewrite algorithm,
the cheapest potential rewrite for target $W_3$ is composed of 
a rewrite for $n_1$ using $a$ of cost 1, 
a rewrite for $n_2$ using $c$ of cost 2,
and the node $n_3$ of cost 2, having a total \ocost{} of 5, 
whereas the next cheapest possible rewrite of $n_3$ uses $b$ and has an 
\ocost{} of 6.
%$\FindNextMinTarget$ chooses $a$ as the first candidate view to $\Refine{}$ 
%within $a+c+n_3$ because the $\ocost{}$ of $a$ (=1) is less than $c$ (=2).
$\FindNextMinTarget$ identifies the potential rewrite composed by $a$, $c$, $n_3$, 
and it chooses $a$ as the first candidate view to $\Refine{}$ within this rewrite, 
because the $\ocost{}$ of $a$ (=1) is less than $c$ (=2).
After refining $a$, the actual $\cost{}$ of $a$ is found to be 4, which is shown 
as a label inside the triangle.
Note this cost is not known ahead of time, which is why the view is originally 
at cost of 1.
Therefore $\minCost{}_1$ is set to 4.
Now since the best known rewrite for 
$W_3$ has a total cost of 11, since $n_1$=4 (using a) + $n_2$=5 + $n_3$=2, then
the value $\minCost{}_3$ is updated from 13 to 11 by $\PropagateBestRewrites{}$. 

Next it attempts the rewrite for $n_3$ using $b$ with an $\ocost{}$ of 6, 
which is less than the next best choice of $(d=3)$+$(c=2)$+$(n_3=2)$
with a total $\ocost{}$ of 7.
The \cost{} of the rewrite of $n_3$ using $b$ was found to be 12 as indicated 
in Figure~\ref{fig-optcost-algo-illustration}.
Since $\minCost{}_3$ is already 11, it is not updated.
The next best choice for $n_3$ is therefore the rewrite with the $\ocost{}$ of 7.
Within that rewrite, $\FindNextMinTarget{}$ chooses to refine $c$,
yielding a rewrite for $n_2$ whose actual cost is 4,
so $\minCost{}_2$ is set to 4.
Then $\minCost_3$ is set to 
$n_1$=4 (using $a$) + $n_2$=4 (using $c$) + $n_3$=2,
with a total cost of 10.
%Figure~\ref{fig-optcost-algo-illustration} captures
%this present state of the algorithm after the first three rewrite attempts.

The algorithm proceeds to the next best possible rewrite of
$(d=3)$+$(c=4)$+$(n_3=2)$ with \ocost{} of 9 which is still better 
than the best known rewrite of $n_3$ with cost of 10.
The algorithm terminates when there are no possible rewrites remaining for 
$W_n$ with a \ocost{} less than $\minCost_n$.
%hence the algorithm cannot stop until it is examined.
%Finally, $e$ can be pruned as it exceeds the current $\minCost{}$ at $n_3$.
Any view with an \ocost{} greater than their target node's $\minCost${} can be pruned away, 
e.g., $e$ at $n_3$ (since $11>10$) and $f$ at $n_1$ (since $5>4$).
\end{example}

It is noteworthy in Example~\ref{ex-algo-description} that had the algorithm
started at $n_3$ first, it would have examined all the candidate
views of $n_3$, resulting in a larger search space than necessary.

\subsection{Proof of Correctness and Work-efficiency}\label{sec-wf-proof}

The following theorem provides the proof of correctness and 
the work-efficiency property of our $\WF{}$ algorithm.
%Due to lack of space, the proof of Theorem~\ref{thm-work-efficient}
% is provided in the extended version~\cite{Lefe2013}.

\begin{theorem}\label{thm-work-efficient}
$\WF{}$ finds the optimal rewrite $r^*$ of $W$ and is work-efficient.
\end{theorem}

\begin{proof}

To ensure correctness, $\WF{}$ must not terminate before finding $r^*$.
Correctness requires that we show the algorithm examines every 
candidate view with $\ocost{}$ less than or equal to $\cost(r^*)$.
To ensure work-efficiency, $\WF{}$ should not examine any extra views 
that cannot be included in $r^*$.
Work-efficiency requires that we show the algorithm must not examine any candidate view 
with $\ocost{}$ greater than $\cost(r^*)$.
We first prove these two properties of $\WF{}$ for a query $W$ containing
 a single target, then extend these results to the case when $W$ contains $n$ targets.

For $W$ with a single target (i.e., $n=1$), proof by contradiction proceeds as follows.
Consider two different rewrites $r$ and $r^*$ such that $r^*$ is the optimal rewrite and $\cost(r^*) < \cost(r)$.
Assume a candidate view $v^*$ produces the optimal rewrite $r^*$.  
Assume another candidate view $v$ produces the rewrite $r$.
Suppose that $\WF{}$ examines $v$, sets $\minCost{}_n$ to $\cost(r)$, and then terminates before examining $v^*$.  
Hence $\WF{}$ incorrectly reports $r$ as the optimal rewrite even though $\cost(r^*) < \cost(r)$. 
Because $\WF{}$ examines candidate views by increasing $\ocost$, then since $v$ was examined before $v^*$, 
it must have been the case that $\ocost(v) < \ocost(v^*)$.
By design, $\WF{}$ will continue until all candidate views whose $\ocost{}$ is less than or equal
to $\minCost{}_n$ have been found.
Given the lower bound property of $\ocost{}$ with respect to $\cost{}$, we have that:
$\ocost(v) < \ocost(v^*) \leq \cost(r^*) < \cost(r) = \minCost_n.$
From the above inequality, it is clear that the algorithm must have examined $v^*$ (and consequently found $r^*$) 
before terminating. %which results in a contradiction.
This results in a contradiction since we assumed earlier that $\WF{}$ terminated before examining $v^*$.
% The correctness of the algorithm is shown since it examines every candidate view with $\ocost{}$ less than or equal to $\cost(r^*)$.

We can similarly prove work-efficiency by contradiction as follows.
Assume that $\ocost(v) > \cost(r^*)$  and as above the algorithm examines $v$ before $v^*$.
This results in the following inequality.
$ \cost(r^*) < \ocost(v) \leq \cost(r) = \minCost_n < \ocost(v^*).$
The results in a contradiction since $\cost(r^*)$ cannot be less than $\ocost(v^*)$, based on the lower-bound property of $\ocost$.
% work-efficiency is shown since it never examines any candidate view with $\ocost{}$ greater than $\cost(r^*)$.

Now we extend this result to $W$ with multiple targets, $n>1$.
It is sufficient to show that the $\WF{}$ algorithm works by reducing $n$ individual search 
problems into a single global search problem that finds the optimal rewrite for the target $W_n$.
Recall that $\WF{}$ instantiates a priority queue $PQ$ at each of the $n$ targets in the plan, where the candidate views
at each $PQ$  are ordered by increasing $\ocost{}$. 
Next we show that the search process degenerates these $n$ PQs into a single virtual global $PQ$  whose elements 
are potential rewrites of $W_n$, ordered by their increasing $\ocost{}$.
%Recall that for each target $W_i$, FindnextMinTarget recursively identifies the lowest cost potential rewrite
Recall that every invocation of $\FindNextMinTarget{}$ identifies the next best target to refine out of all targets in $W$.
It composes the lowest cost potential rewrite for $W_n$ by recursively visiting each target $W_i$ and selecting the cheaper among
 either the candidate view at the front of the $PQ$ for $W_i$ or the current best known rewrite for $W_i$. 
Thus this recursive procedure identifies the current lowest cost potential rewrite of $W_n$, 
in effect gradually exploring the space of potential rewrites of $W_n$ by their increasing $\ocost{}$.
This creates a virtual global $PQ$ whose elements are potential rewrites of $W_n$ and is ordered by their $\ocost{}$.
% Comment:  of that curret best potential rewrite found by the recursive procedure of FindNextMinTarget,
% of all the candidate views that compose thi rewrite, FindNextMinTarget actually returns the view with the lowest ocost.

For example, in Figure~\ref{fig-optcost-algo-illustration}, note that the first potential rewrite for $W_n$ 
is composed of $a,c,n_3$ with an $\ocost{}$ of 5, 
while the next potential rewrite of $W_n$ is composed of $b$ with an $\ocost{}$ of 6.  
Furthermore, notice that the rewrite produced by $a,c, n_3$ has a cost of 10, which is always greater than or 
equal to its corresponding $\ocost{}$ due to the lower-bound property.  
We have now reduced $n$ priority queues of candidate views ordered by $\ocost{}$ to a single global priority 
queue of potential rewrites of $W_n$ ordered by $\ocost{}$.  
This completes the proof since we already showed above that $\WF{}$ with a single $PQ$ ordered by $\ocost{}$ 
will find the optimal rewrite and is work-efficient.
\end{proof}

\section{ViewFinder}\label{sec-viewfinder}

The key feature of $\VF{}$ is its $\ocost{}$ functionality
that enables it to incrementally explore the the space of rewrites using
the views in $V$.
%This functionality is used by $\WF{}$ to explore 
%which is used by the $\WF{}$ to explore
%This allows it to prune unnecessary sub-spaces as shown in 
%Section~\ref{sec-solution-outline}, and visit
% Thus we incorporate $\ocost{}$ with the traditional approach
% which consists of finding {\em complete} candidate views and enumerating
% rewrites for each candidate.
As noted earlier in Section~\ref{sec-equivalence}, rewriting queries 
using views is known to be a hard problem.
Traditionally, methods for rewriting queries using views 
for SPJG queries use a two stage approach~\cite{Hale2001, Agra2000}.
The pruning stage determines which views are \emph{relevant} to the query,
and among the relevant views those that contain
all the required join predicates  are termed as \emph{complete},
otherwise they are called {\em partial} solutions.
This is typically followed by a merge stage that joins the partial
solutions using all possible equijoin methods on all join orders to form
additional relevant views.
The algorithm repeats until only those views that are useful for answering 
the query remain.

We take a similar approach in that we identify partial and
complete solutions, then follow with a merge phase.
The $\VF{}$ considers candidate views $C$ when 
searching for rewrite of a target.
$C$ includes views in $V$ as well as views
formed by ``merging'' views in $V$ using a
\textsc{Merge} function, which is an implementation of a 
standard view-merging procedure (e.g.,~\cite{Agra2000,Hale2001}).
Traditional approaches merge partial solutions to create complete solutions,
continuing 
until no partial solutions remain.
This ``explodes'' the space of candidate views exponentially up-front.
In contrast, our approach gradually explodes the space, 
resulting in far fewer candidates views from being considered.

% Here we will describe the functionality of the \VF{}, and later in
% Section~\ref{sec-best-first-workflow-rewrite} we
% will present the \WF{} which exploits the \VF's{} ability to incrementally examine
% the space of solutions for each target.
% However, the direct applicability of existing approach
Additionally, with no early termination condition, existing approaches would need to explore 
the space exhaustively at all targets. 
%Thus we desire a rewriting algorithm  that can enumerate the space and 
%incrementally explore only as much as required, frequently stopping and 
%resuming the search as requested by \WF.
The $\VF{}$ incrementally grows and explores only as much
of the space as needed, frequently stopping and resuming the search as
requested by \WF.

% We note that while an equivalent rewrite for a target may exist,
% the \VF may never be asked to find it,
% as illustrated by Example~\ref{example-greedy-plumbing} for
% the case of the candidate view with cost 12 at $n_3$. 

%\subsection{V\fakesc{iew}F\fakesc{inder} Algorithm}
\subsection{The $\VF{}$ Algorithm}

The $\VF{}$ is presented in Algorithm~\ref{alg-viewfinder}.
There is an instance of $\VF{}$ instantiated at each target, which is 
{\em stateful}; enabling it to start, stop, and resume the incremental 
searches at each target.
The $\VF{}$ maintains state using a priority queue ($\PQ$) of candidate views,
ordered by \ocost. 
% for each target.
$\VF{}$ implements the {\sc Init}, {\sc Peek}, and {\sc Refine} functions.
% which we describe next.

The \Init{} function instantiates an instance of the $\VF{}$ 
with a query $q$ representing a target $W_i$, 
and the set of all materialized views $V$ is added to $\PQ$.

The $\Peek{}$ function is used by $\WF{}$ to obtain the $\ocost{}$ of the head 
item in a $\PQ$.

The $\Refine{}$ function is invoked when $\WF{}$ asks the $\VF{}$
to examine the next candidate view.
At this stage, the $\VF{}$ pops the head item $v$ out of $\PQ$.
The $\VF{}$ then generates a set of new candidate views $M$
by merging $v$ with previously popped  candidate views (i.e., views in $Seen$),
thereby incrementally exploding the space of candidate views.
Note that $Seen$ only contains candidate views that have an $\ocost{}$ 
less than or equal to that of $v$.
Merged views in $M$ are only retained if they are not already in $Seen$.
Then views in $M$ are inserted into $\PQ$ and $v$ is added to $Seen$.
% TODO: Need a proof for that statement

A property of $\ocost{}$ (provided as a theorem below) is that the candidate views in $M$ have an
$\ocost{}$ that is greater than that of $v$ and hence none
of these views should have been examined before $v$.
Critically, this enables \VF to perform a gradual explosion of the space of 
candidate views.
At this point, the view is considered for a rewrite as described next.

\begin{theorem}
The $\ocost{}$ of every candidate view in $M$ that is not in $Seen$
is greater than or equal to the $\ocost{}$ of $v$.
\end{theorem}
\begin{proof}
The proof sketch is as follows. The theorem is trivially true for 
$v \in V$ as all candidate views in $M$ cannot be in $Seen$
and have $\ocost$ greater than $v$. 
If $v \notin V$, it is sufficient to point out
that all constituent views of $v$ are already in $Seen$
since they must have had $\ocost{}$ lesser or equal to $v$
Hence all candidate views in $M$ with $\ocost{}$ smaller than $v$ are
 already in $Seen$, and those with $\ocost{}$ greater than $v$ will be added to $PQ$
if they are not already in $PQ$.
 %all the 
%candidate views in $M$ are already present in $Seen$ if smaller, or if greater,
%are still in $\PQ$.
\end{proof}

\begin{algorithm}[h!]
\caption{$\VF$\label{alg-viewfinder}}
{\small
\begin{algorithmic}[1]
\Function{\sc Init}{$query$, $V$}\label{alg-vf-init}
 \State Priority Queue $PQ$ \qlet $\emptyset$; $Seen$ \qlet $\emptyset$; Query $q$ 
  \State $q$ \qlet $query$
  \ForAll{$v \in V$}
     \State $PQ$.\texttt{add}($v$, $\ocost(q,v))$
  \EndFor
\EndFunction
\end{algorithmic}
\Line
\begin{algorithmic}[1]
\Function{\sc Peek}{}\label{alg-vf-peek}
%\State \Return ($PQ$.\texttt{peek}().$\ocost$: $\infty$) ? 
\State {\bf if} $PQ$ is not empty {\bf return} $PQ$.\texttt{peek}().$\ocost$ {\bf else} $\infty$
%   \If{not $PQ$.\texttt{empty}()}
%   \State \Return $PQ$.\texttt{peek}().$\ocost$ 
%   \Else 
%   \State \Return $\infty$
%   \EndIf  
\EndFunction
\end{algorithmic}
\Line
\begin{algorithmic}[1]
\Function{\sc Refine}{}\label{alg-vf-refine}
  \If{\texttt{not} $PQ$.\texttt{empty}()}
    \State $v$ \qlet $PQ$.\texttt{pop}()
    \State $M$ \qlet \merge($v$, $Seen$) \Comment{Discard from $M$ those in $Seen \cap M$}
    \ForAll{$v' \in M$}
      \State $PQ$.\texttt{add}($v'$, $\ocost(q,v')$)
    \EndFor
    \State $Seen$.\texttt{add}($v$)
    \If{\complete($q$, $v$)}
      \State \Return $\RE(q, v)$
    \EndIf
  \EndIf \\
  \Return {\sc null}
\EndFunction
\end{algorithmic}}
\end{algorithm}

\subsection{Rewrite Enumeration}\label{sec-rewrite-enumeration}

Given the computational cost of finding valid rewrites,
$\WF{}$ limits the invocation of the $\RE{}$ algorithm using two strategies.
% we employ the following two strategies.
First,  we avoid having to apply $\RE{}$ on every candidate view by using \complete.
Second, we delay the application of $\RE$ to every complete view 
by determining a lower bound on the cost of a rewrite by using \ocost.
% $v$ should one exist.
%For the lower bound we use the $\ocost$, which helps   
% The goal of guessing completeness and estimating a lower bound is
%to quickly distinguish between potentially low cost rewrites from those with 
%higher cost.

% Suppose that a candidate view $v$ is guessed to be complete.
% To determine if an equivalent rewrite exists requires finding at
% least one valid rewrite among all possible rewrites,
% is known to be a hard problem~\cite{Levy1995}.
The \RE{} procedure (pseudo-code not shown but described here) searches for a 
valid rewrite of a query $q$ using a view $v$ that is guessed to be complete.
% \RE{} takes as input a query $q$ and a view $v$ that is guessed to be
% complete and outputs a rewrite $r$ if one exists.
Given that \complete{} can result in false positives, a rewrite may not be 
found.
If a rewrite is found, \RE{} returns the rewrite and its cost as determined 
by the \cost{} function.
%which corresponds to the cheapest execution \textit{plan} that implements the 
%rewrite.
%Equivalence is determined by ensuring that the rewrite and query contain the 
%same attributes, filters, and group-by as detailed in 
%Section~\ref{sec-equivalence}.
%This type of equivalence testing is simple but effective because we produce 
% all possible rewrites using $v$, and any perceived shortcoming with the test 
%can be overcome by extending  $L_R$. Although this type of equivalence 
% testing may seem simple it is effective because we produce all possible rewrites%using $v$, and any perceived shortcoming with the test can be overcome by extending  $L_R$.
%We replace an SPJ+G expression with a single view.  as in GL2001

From Section~\ref{sec-problem-definition}, recall that the rewrite process only considers 
a subset of the UDFs in the system and standard relational operators SPJGA.
These are the only \textit{transformations} considered by \RE.
The rewrite process searches for equivalent rewrites of $q$ by applying {\em compensations}~\cite{Zaha2000}
to a view $v$ that is guessed to be complete for $q$, using only the transformations. 
\RE{} does this by generating all permutations of required compensations and
testing for equivalence, which
amounts to a brute force enumeration of all possible rewrites that can be produced by applying the transformations. 
This makes the case for the system to keep the set of transformations small since
this search process is exponential in the size of this set.
When the transformations are restricted to a fixed known set, 
it may suffice to examine a polynomial number of rewrites attempts, as in~\cite{Grum2003} 
for the specific case of simple aggregations involving group-bys.
Such approaches are not applicable to our case as 
the system has the flexibility to add any UDF to the set of transformations.

%% JL

%% JL temp

\section{Experimental Evaluation}\label{sec-experimental-evaluation}

\twofigrow{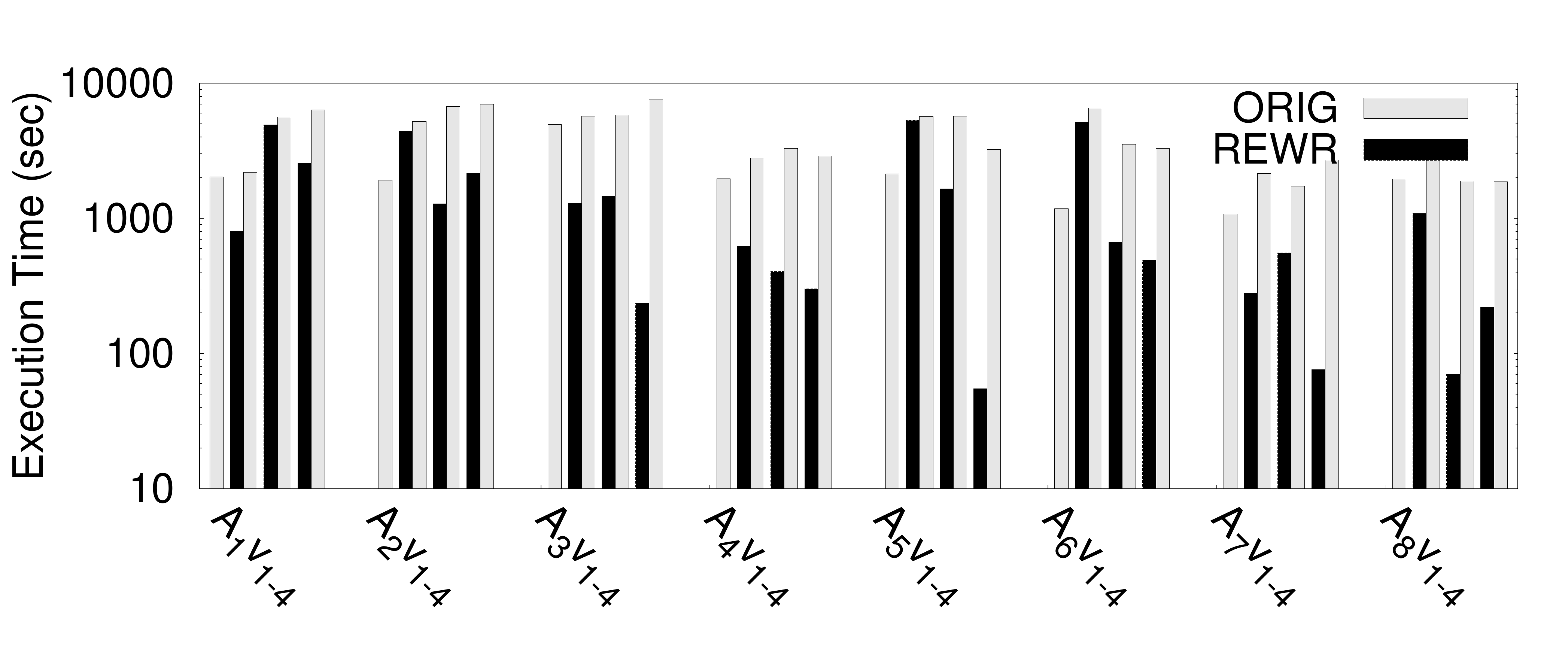}
{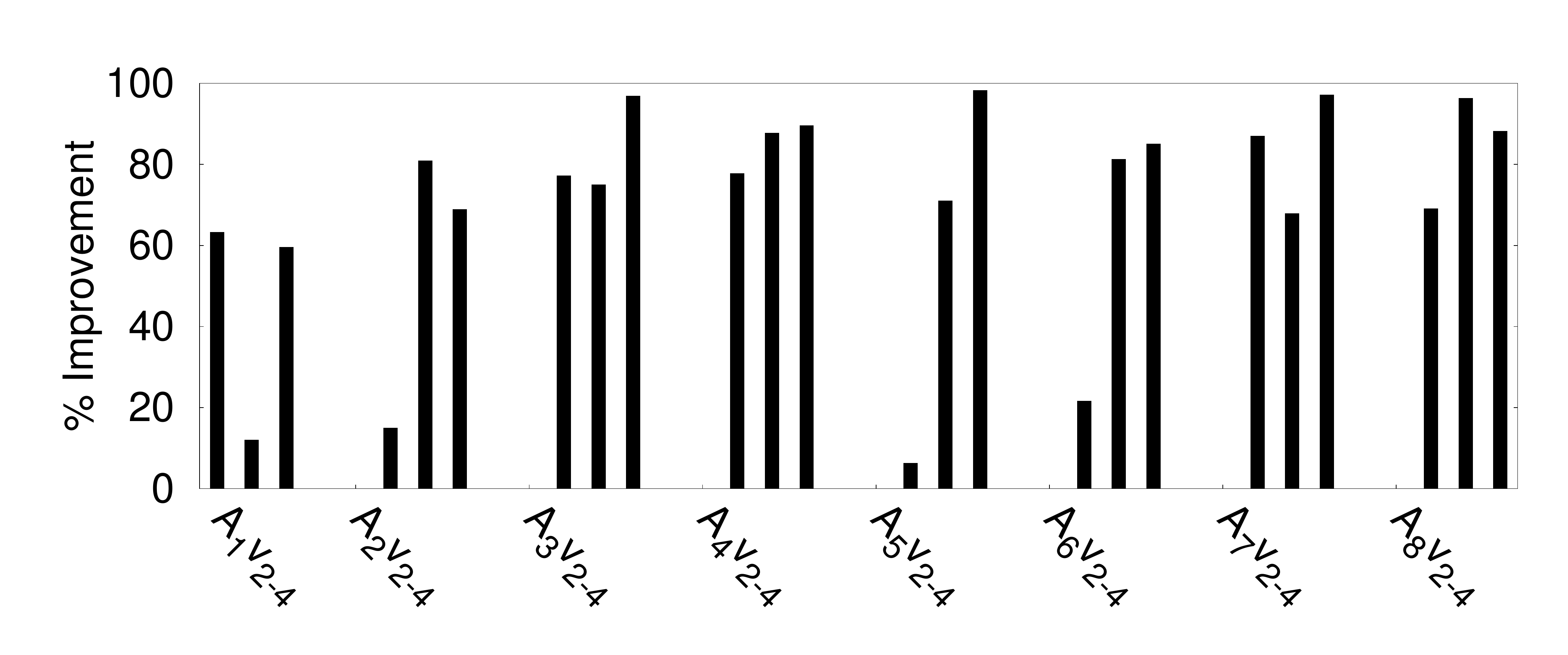}
{Query evolution comparisons for (a) execution time (log-scale),
and (b) execution time improvement.}{3.65in}

In this section, we present an experimental study we conducted in order to 
validate the effectiveness of $\WF{}$ in finding low-cost rewrites of complex queries. 
We first evaluate our methods in two scenarios.
The {\em query evolution} scenario (Section~\ref{sec-query-evolution}) represents a user 
iteratively refining a query within a single session.
This scenario evaluates the benefit that each new query version can receive from the opportunistic
views created by previous versions of the query.
The {\em user evolution} scenario (Section~\ref{sec-experiments-ue}) represents a new user
  entering the system presenting a new query.
This scenario evaluates the benefit a new query can receive from the opportunistic
views created by queries of other ``similar'' users.
We compare the performance of our algorithm with a competing approach in (Section~\ref{sec-scalability}).
%with a baseline dynamic programming approach
%that searches at all targets without using $\ocost{}$.
Next, we evaluate the scalability  (Section~\ref{sec-scalability}) of our rewrite algorithm in comparison
to the dynamic programming approach.
We then compare our method to cache-based methods (Section~\ref{sec-experiments-restore}) 
 that can only reuse identical previous results.
Lastly, we show the performance of our method (Section~\ref{sec-experiments-storage-reclamation}) under a storage reclamation policy that drops opportunistic views.
% Lastly, as a sanity check (Section~\ref{sec-experiments-dbx}) we compare the quality of rewrites produced by our algorithm with a state-of-the-art  DBMS.

\subsection{Methodology}\label{sec-evaluation-setup}
% UDF is written in Hive, perl, python, MR but the transformation applied
% by the UDF is expressed in terms of a grammar that describes how an input
% attributes, filter and groupby properties are modified by the UDF.

Our experimental system consists of 20 machines running Hadoop.
We use HiveQL as the declarative query language, and Oozie as a
job coordinator.
The MR UDFs are implemented in Java, Perl, and Python and executed
using the HiveCLI.
UDFs implemented in our system include a log parser/extractor,
text sentiment classifier, sentence tokenizer, lat/lon extractor, 
word count, restaurant menu similarity, and geographical 
tiling, among others.
All UDFs are annotated using the model, as per the example
annotations given in Section~\ref{sec-udf-example}.

Our experiments use the following three real-world datasets totalling over 1TB:
a Twitter log containing 800GB of tweets, 
a Foursquare log containing 250GB of user check-ins, 
and a Landmark log containing 7GB of 5 million landmarks including their locations.
The identity of a social network user (\texttt{user\_id}) is common across
Twitter and Foursquare logs, while the identity of a landmark
 (\texttt{location\_id}) is common across Foursquare and Landmarks logs.

For all experiments, we report on the following metrics.
Experiments on query execution time report both the original execution time 
of the query 
in Hive, labelled as \textsc{orig}, and the execution time of the rewritten 
query, labelled as \textsc{rewr}.
The reported time for \textsc{rewr} includes the time to run the \WF{} 
algorithm, the time to execute the rewritten query, and any time spent on 
statistics collection (Section~\ref{sec-preliminaries}).
Experiments on rewrite algorithm runtime report the total optimization time 
used by the  algorithm to find a rewrite of the original query using the views 
in the system.
For these experiments, \textsc{bfr} denotes  the use of our \WF{} algorithm, and  
\textsc{dp} represents a competing rewrite approach.  
\textsc{dp} does not use \ocost{} and searches exhaustively for rewrites at every target, then 
applies a dynamic programming solution to choose the best subset of rewrites found at each target
to rewrite the query.
It is to be noted that both algorithms produce identical rewrites (i.e., $r^*$).
The primary comparison metric for \textsc{bfr} and \textsc{dp} is the algorithm runtime,
and secondary metrics are the number of candidate views examined during the search for rewrites,
and the number of valid rewrites produced, i.e., the space explored and rewrites attempted
before $r^*$ is found.

\subsection{Query Workload}\label{sec-evauation-query-scenarios}

The experimental workload contains 32 queries simulating 8 analysts $A_1$--$A_8$ who write 
complex analytical queries for business marketing scenarios from~\cite{lefe2013b}.
These queries represent exploratory analysis on big data, and contain UDFs.
%As there is no  available representative workload containing many UDFs that captures 
%exploratory analysis on big data, we developed a workload based on our 
%experience and discussions with analysts from several industries.
%A detailed description of the workload is available in~\cite{lefe2013b}.
%The queries model the analysts' iterative explorations of big data for 
%realistic marketing scenarios involving restaurants, and use social media data.
%We derived these representative queries from extensive conversations with 
%business users on their typical analytics use cases.
 %, which are described in 
 %Table~\ref{tab-experiment-scenarios}.
Each analyst in the workload poses 4 versions of a query, representing the 
initial query followed by three subsequent revisions made during data 
exploration and hypothesis testing.
Hence, there is some overlap expected between subsequent version of a query.
%We use these 32 queries for all of our evaluations.
%Each query version captures the analyst's iterative 
%...attempt  at bringing the query closer to her intended goal.
%refinements to the query during data exploration and hypothesis testing.
The queries are long-running with many operations, and executing 
the original versions of the queries in Hive created 17 opportunistic materialized 
views on average.

As each query has multiple versions, we use $A_iv_j$ to denote Analyst $i$ 
executing version $j$ of her query, and version $j$ represents a revised version of $j-1$.
We briefly describe the first two versions of $A_1$'s query in Example~\ref{ex-query-a2v2}.
The queries reference data from Twitter (TWTR), Foursquare (4SQ), and Landmarks (LAND) data, 
and UDFs are denoted in all caps.  

\begin{example}\label{ex-query-a2v2}
Analyst1 ($A_1$) wants to identify a number of ``wine lovers'' to
send them a coupon for a new wine being introduced in a local region.

Query $A_1v_1$:
(a) From TWTR, apply UDF-CLASSIFY-WINE-SCORE on each user's tweets and group-by user to produce wine-sentiment-score for each user.
Apply a threshold on wine-sentiment-score. % above \underline{$x_1$}.   
(b) From TWTR, compute all pairs $\langle u_1,u_2\rangle$ of users that communicate
with each other, assigning each pair a friendship-strength-score based on the number of times they communicate.
%Select user pairs having friendship-strength-score greater than \underline{value1}.
Apply a threshold on friendship-strength-score. %above \underline{$x_2$}.
(c) From TWTR, apply UDAF-CLASSIFY-AFFLUENT on users and their tweets.
Join results from (a), (b), (c) on user\_id.

Query $A_1v_2$: Revise the previous version by reducing the wine-sentiment-score 
threshold, adding new data sources (4SQ and LAND) to find the check-in counts for 
users that check-in to places of type wine-bar, then threshold on count and 
join with users found in the previous version. 
Query Versions 3 and 4 are revised similarly but omitted here.  
\end{example}
\normalsize

\subsection{Experimental Results}

\threefig{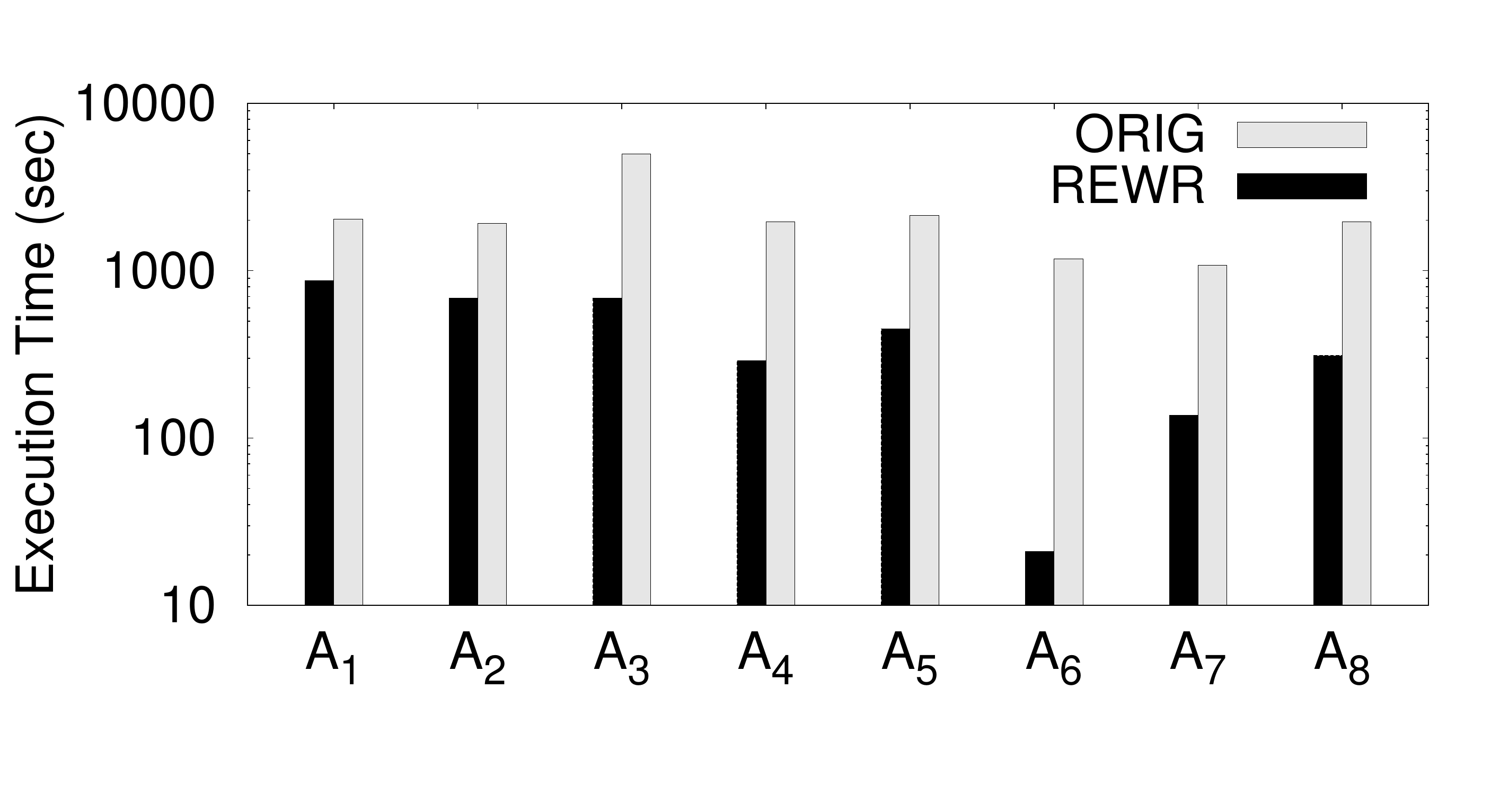}
{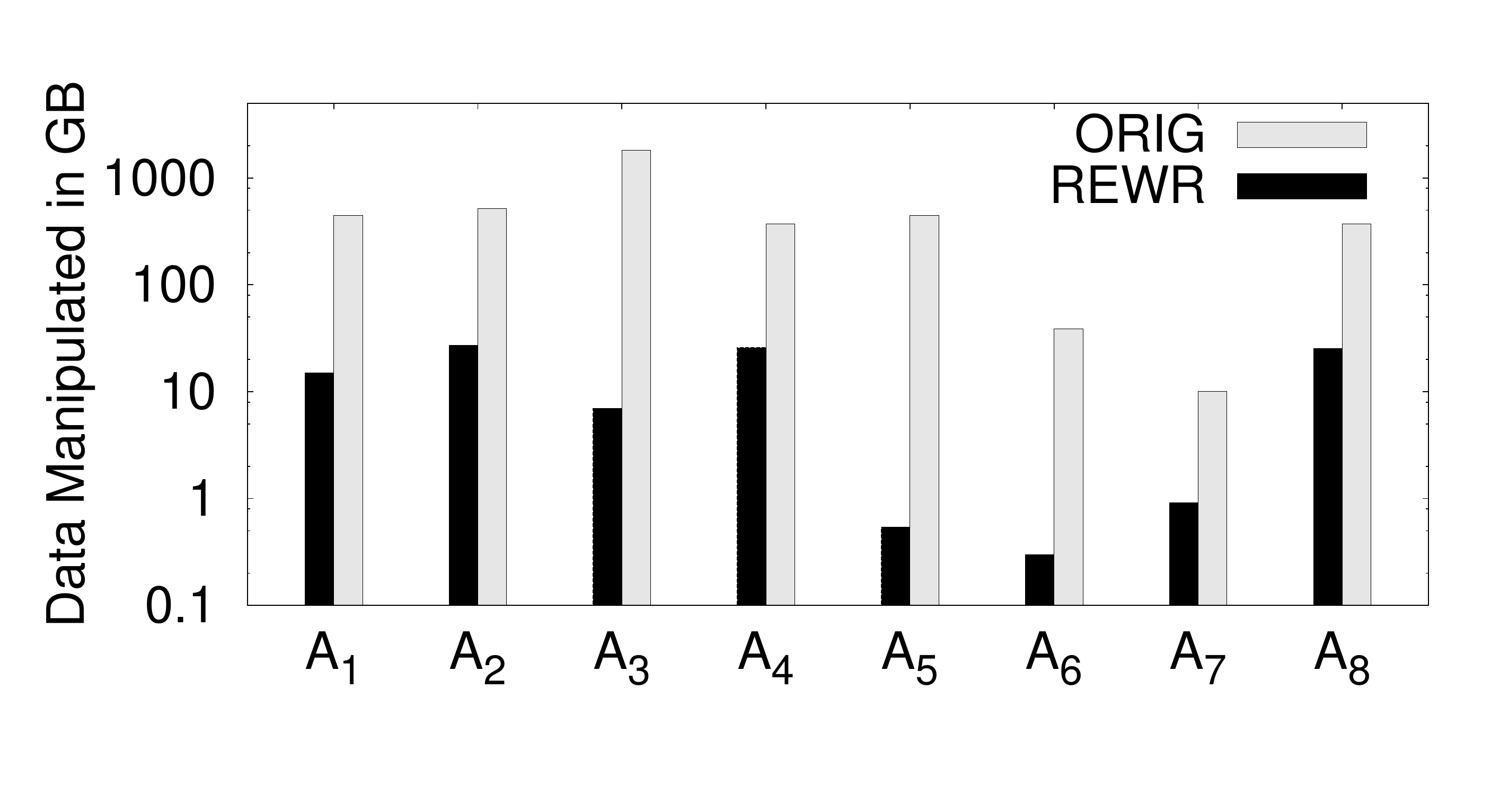}
{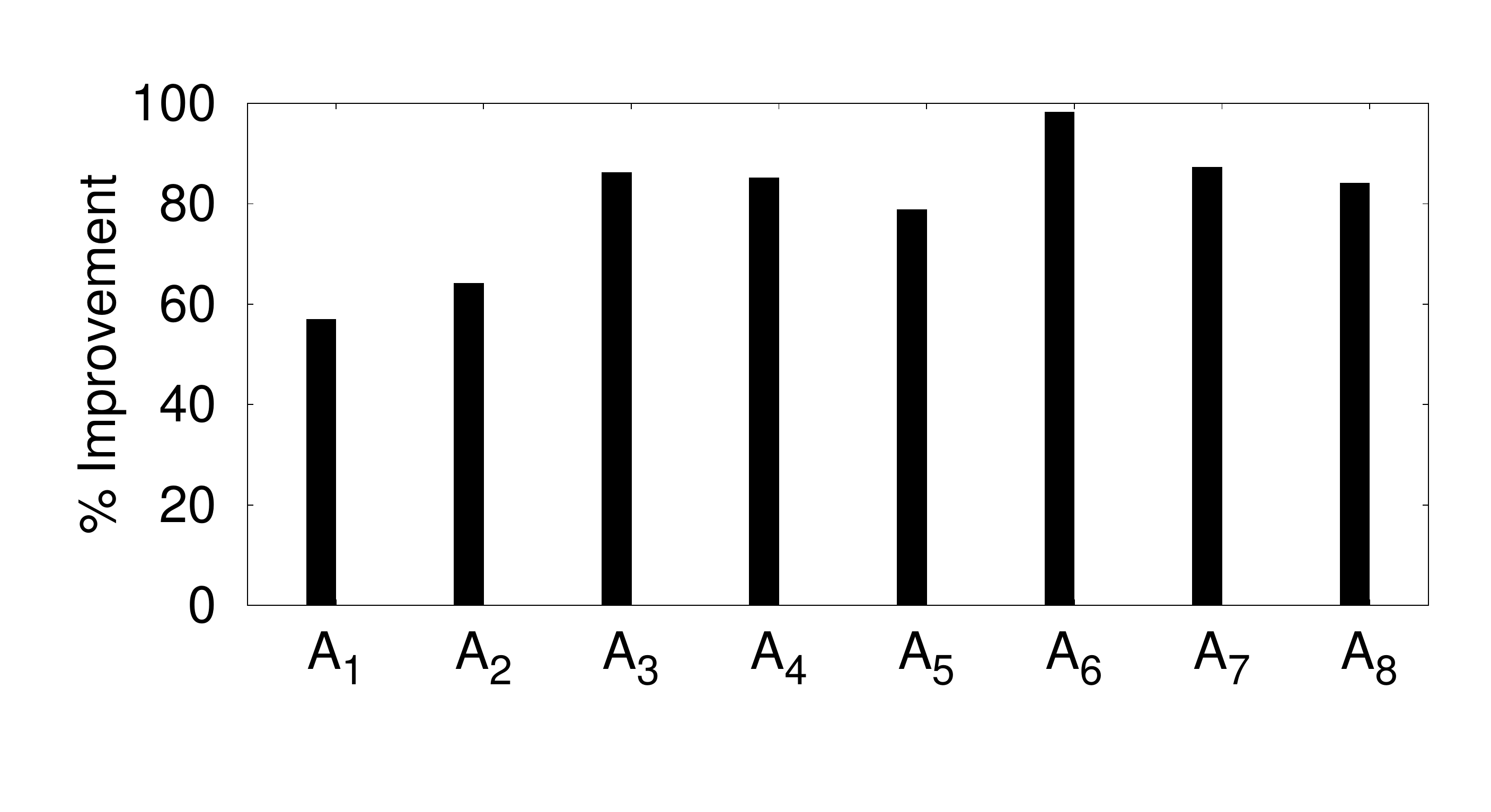}
{User Evolution Comparisons for (a) execution time (log-scale), (b) data moved,
 and (c) execution time improvement.}{2.35in}

\subsubsection{Query Evolution}\label{sec-query-evolution}

%This section presents results for the query evolution scenario.
In this experiment, for each analyst $A_i$, query $A_iv_1$ is executed, 
followed by query $A_iv_2$,  $A_iv_3$, and $A_iv_4$.
Figure~\ref{fig-we-novf-vs-vf-running-time}(a) shows the execution time of 
the original query (\textsc{orig}) and the rewritten query (\textsc{rewr}), 
and  Figure~\ref{fig-we-novf-vs-vf-running-time}(b) reports the 
percent improvement in execution time of \textsc{rewr} over \textsc{orig}.
Figure~\ref{fig-we-novf-vs-vf-running-time}(b) shows \textsc{rewr} provides 
an overall improvement of 10\% to 90\%; 
with an average improvement of 61\% and up to an order of magnitude.
As a concrete data point, $A_5v_4$ requires 54 minutes to execute \textsc{orig},
but only  55 seconds to execute the rewritten query (\textsc{rewr}).
%The amount of data manipulated in Hive (i.e., read/write/shuffle bytes) 
%is not shown here, but closely follows the
%same trend in Figure~\ref{fig-we-novf-vs-vf-running-time}(a).
% Data moved results are not reported here because later we present the results 
% showing similar data movement trends for the user evolution experiments in 
% Section~\ref{sec-experiments-ue}, 
%Figure~\ref{fig-ue-novf-vs-vf-running-time}(b).
\textsc{rewr} has much lower execution time because 
it is able to take advantage of the overlapping nature of the queries, e.g.,  
version 2 has some overlap with version 1.
\textsc{rewr} is able to reuse previous results, which provides a significant 
savings in computation time and data movement (read/write/shuffle) costs.
%These results show the benefit of our methods for the query evolution scenario,
%where a user benefits from opportunistic views created during execution
%of 

\subsubsection{User Evolution}\label{sec-experiments-ue}
In this experiment, we first execute the first version of each analyst's 
query except one (holdout analyst  $A_i$).
Then, we execute the first version of the holdout analyst's query ($A_iv_1$), and repeat this with a different
holdout analyst each time.  
Figure~\ref{fig-ue-novf-vs-vf-running-time}(a) shows the execution time for \textsc{rewr} and \textsc{orig}
for each different holdout analyst along the x-axis, while 
Figure~\ref{fig-ue-novf-vs-vf-running-time}(b) shows the 
corresponding data manipulated (read/write/shuffle) in GB.
These results demonstrate that the execution time is always lower 
for \textsc{rewr}, with the amount of data moved showing a similar trend.
The percentage improvement in execution time is given in 
Figure~\ref{fig-ue-novf-vs-vf-running-time}(c) which shows \textsc{rewr}
results in an overall improvement of about 50\%--90\%.
This scenario mimics a new analyst arriving in a system, and the results show
the benefit that is obtained by reusing previous results from many other analysts that 
pose queries on the same data.
Of course, these results are workload dependent but they show that even when 
analysts query the same data sets while testing different hypothesis,
our approach is able to find some overlap and benefit from previous results.

\begin{table}[!h] \vspace{-0.2in}
\caption{\label{fig-se-novf-vs-vf-time-improvement}
Improvement in execution time of $A_5v_3$ as more
analysts become present in the system.}
\centering
\begin{tabular}{| p{2.45cm} | p{0.35cm}| p{0.45cm} | p{0.45cm}| p{0.45cm} | p{0.45cm}|  p{0.45cm} | p{0.45cm}|}
\hline
Analysts added & 1 & 2 & 3 & 4 & 5 & 6 & 7 \\ \hline
Improvement    & 0\% & 73\% & 73\% & 75\% & 89\% & 89\% & 89\% \\ \hline
\end{tabular}
\vspace{-0.1in}
\end{table}

As an additional experiment for user evolution, we first execute a single analyst's query ($A_5v_3$) 
with no opportunistic views in the system, to create a baseline execution time.
Then we ``add'' another analyst by executing all four versions of that
analyst's query, which creates new opportunistic views. 
Then we re-execute $A_5v_3$ and report the execution time 
improvement over the baseline, 
and repeat this process for the other remaining analysts.
We chose $A_5v_3$ as it is a complex query that uses all three logs. 
Table~\ref{fig-se-novf-vs-vf-time-improvement} reports the 
execution time improvement after each analyst is added, 
showing that the improvement increases when more opportunistic views 
are present in the system.
%The benefit is obviously dependent on similarity to other
%analysts' queries.

\subsubsection{Algorithm Comparisons}\label{sec-scalability}

\threefig{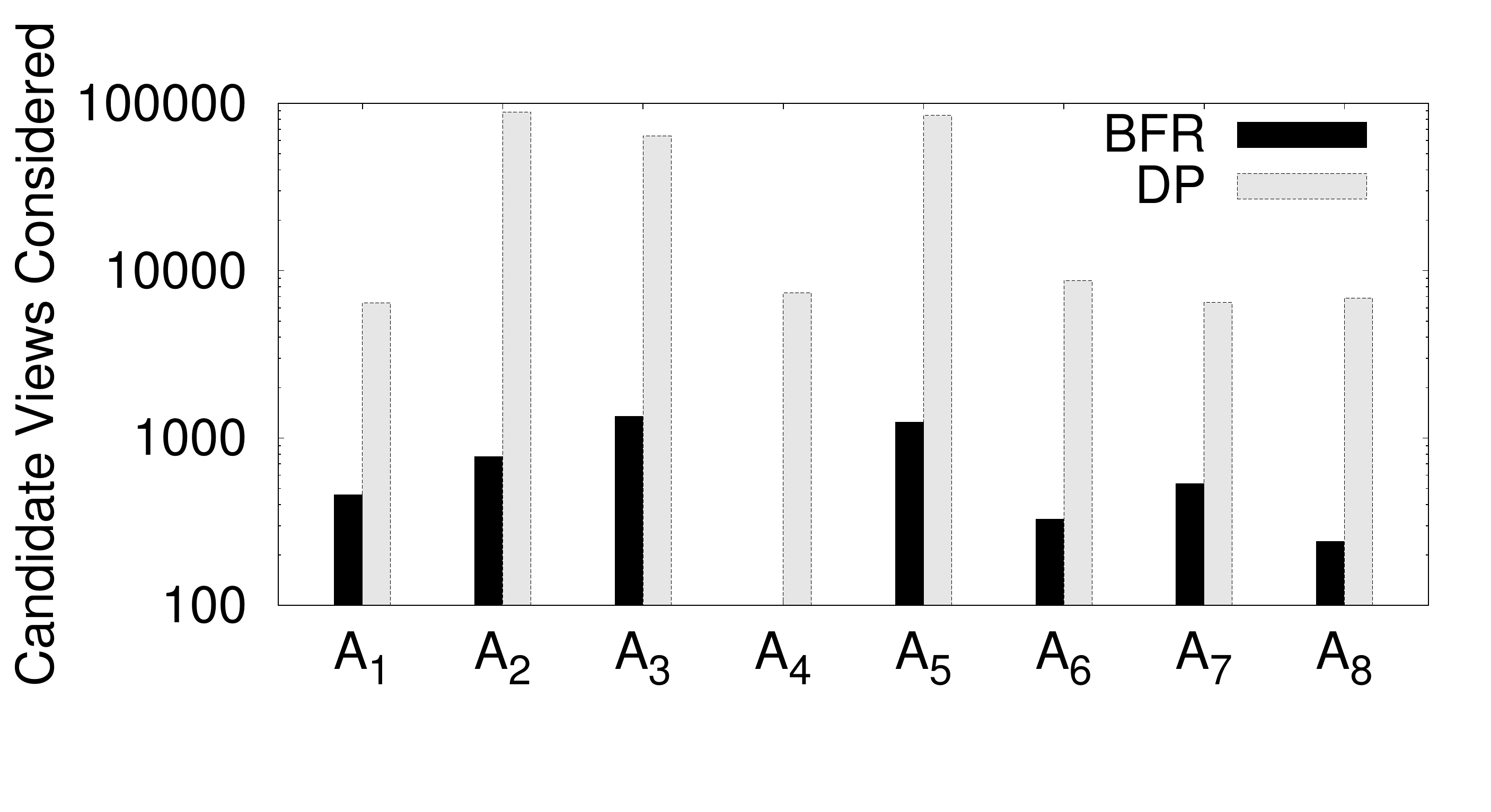}
{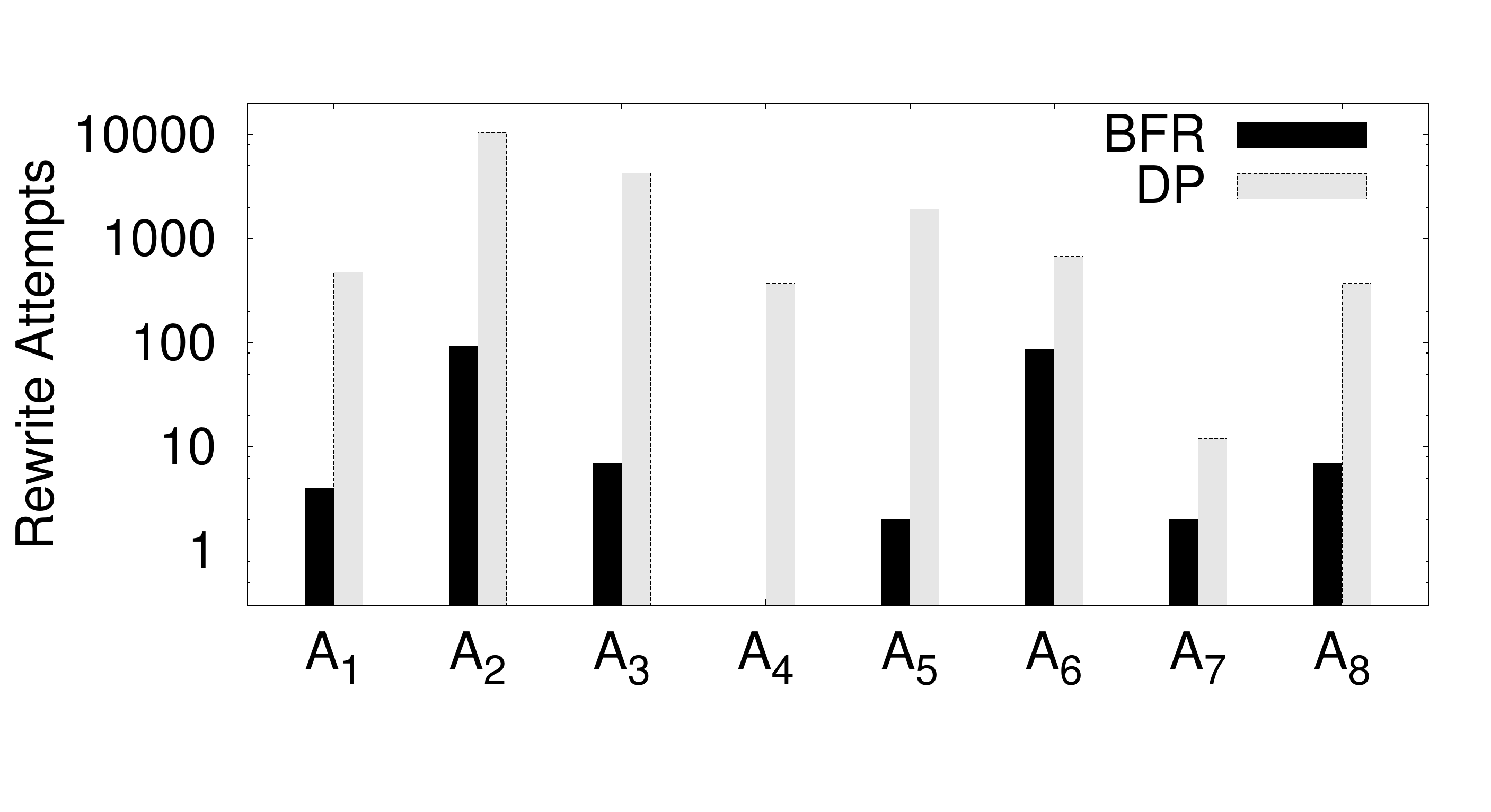}
{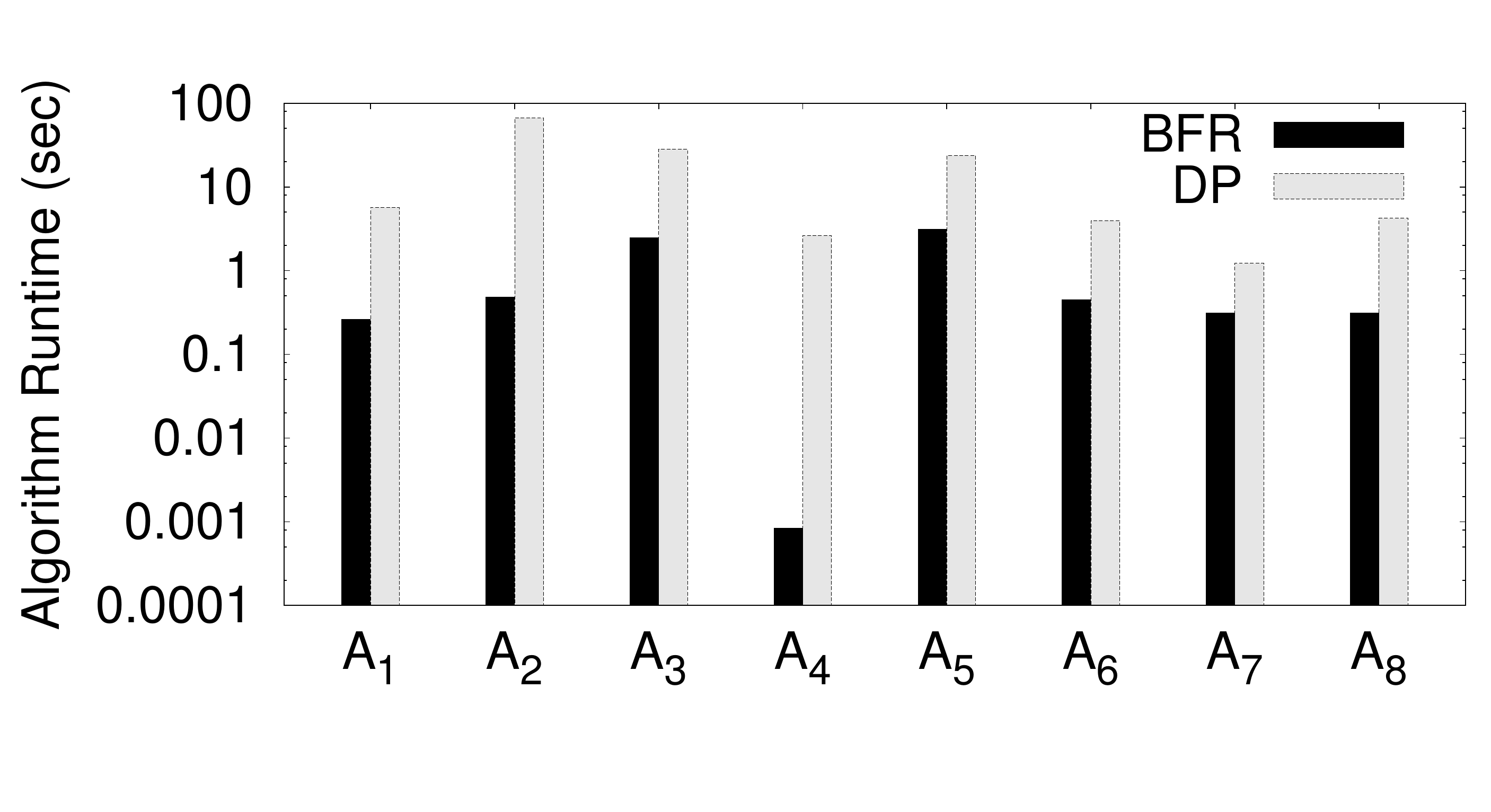}
{Algorithm Comparisons for (a) candidate views considered, (b) refine 
attempts, and  (c) Algorithm runtime (log-scale).}{2.35in}

In the first experiment we compare \textsc{bfr} to \textsc{dp}  in terms of the 
number of candidate views considered, the number of times the algorithm attempts
a rewrite, and the algorithm runtime in seconds.
We use the user evolution scenario from the previous experiment, where 
there were approximately 100 views in the system when each holdout analyst's query
was executed.
Figure~\ref{fig-ue-bfs-vs-dp-candidates-explored}(a) shows that even though both 
algorithms find identical rewrites, \textsc{bfr} searches much less of the space than \textsc{dp}
since it considers far fewer candidate views when searching for rewrites.
Similarly, Figure~\ref{fig-ue-bfs-vs-dp-candidates-explored}(b) shows that \textsc{bfr} 
attempts far fewer rewrites compared to \textsc{dp}. 
This improvement can be attributed to \complete{} identifying the promising candidate views,
and \ocost{} enabling  \textsc{bfr} to incrementally explore the candidate views, thus applying 
\RE{} far fewer times.
Together, these contribute to \textsc{bfr} doing far less work than \textsc{dp}, which 
is reflected in the algorithm runtime shown in Figure~\ref{fig-ue-bfs-vs-dp-candidates-explored}(c).
This shows our algorithm results in significant savings
due to the way it controls the exponential burst of candidate views,  
growing the candidates set incrementally as needed, and by controlling the 
number of times it attempts a rewrites; all of which are all computationally 
expensive since they are exponential in either the number of candidate views 
or the number of transformations.
Next we show how \textsc{bfr} scales as the number of views increases.

%We control the burst of candidates, exponential in the number of cands. 
%We are smart about what we attempt to rewrite, since this is a very expensive process, 
%also exponential.
%Next we show how this helps us to scale better than \textsc{dp}.

%The number of rewrite attempts corresponds to the candidate views examined by
% $\RE{}$ when searching for a valid rewrite with low cost.
%As noted previously this process is computationally expensive.
%These results show for many cases it is possible for $\WF{}$ 
%to find $r^*$ by considering far fewer candidate views as shown in 
%Figure~\ref{fig-ue-bfs-vs-dp-candidates-explored}(a), 
%and orders of magnitude fewer rewrites attempted as shown in 
%Figure~\ref{fig-ue-bfs-vs-dp-candidates-explored}(b),
%resulting in significant savings in algorithm  runtime as shown in 
%Figure~\ref{fig-ue-bfs-vs-dp-candidates-explored}(c).
%This savings is due to \WF{}'s use of $\ocost{}$ that enables
%considering the promising candidate views earlier in the search,
% and exploding the space of candidate views incrementally as needed.
% This shows our algorithm results in a significant reduction
% in the rewrite search space.

\onefig{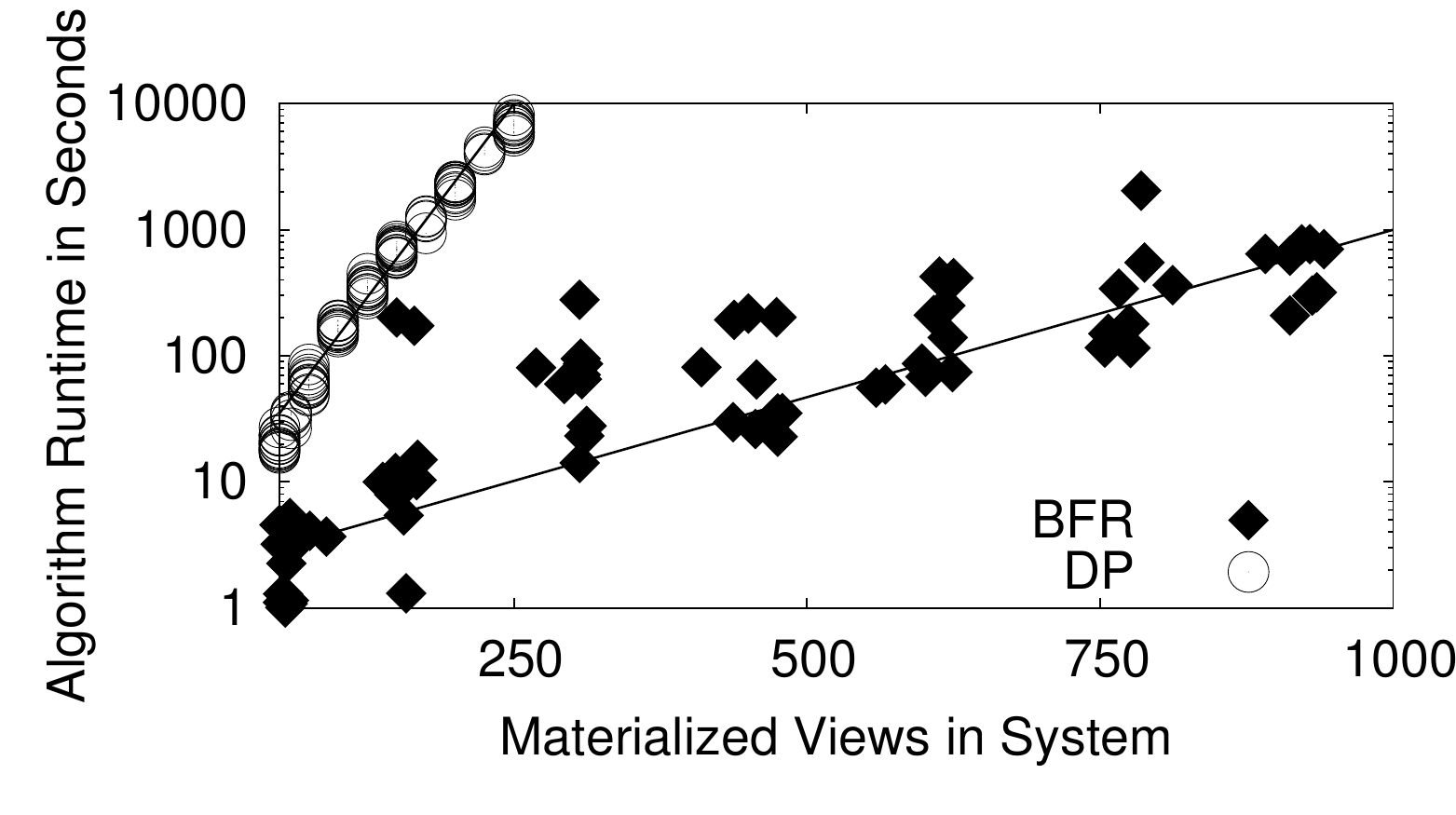}
{Runtime of \textsc{bfr} and \textsc{dp} for varying number of MVs.}
{2.25in}

%Next we evaluate the algorithm runtime as we vary the number of views in 
% the system from 1--1000 to show how the algorithms scale with an increasing 
% number of views.

In the second experiment, we test the scalability of the algorithms by scaling
 up the number of views in the system from 1--1000 and report the algorithm 
runtime for both \textsc{bfr} and \textsc{dp} as they search for rewrites for 
one query ($A_3v_1$).
During the course of design and development of our system,  we created and
retained about 9,600 views; from these we discarded duplicate views as well 
as those views that are an exact match to the query 
(simply to prevent the algorithms from terminating trivially).  
In Figure~\ref{fig-mv-vary-run-time}, the x-axis reports the the number of 
views (views are randomly drawn from among these candidates), while the y-axis 
reports the algorithm run time (log-scale).
\textsc{dp} becomes prohibitively expensive even when 250 MVs
are present in the system, due to the exponential search space.
\textsc{bfr} on the other hand scales much better than \textsc{dpr} and has
a runtime under 1000 seconds even when the system has 1000  views relevant
to the given query.
This is due to the ability of \textsc{bfr} to control the exponential space 
explosion and incrementally search the rewrite space in a principled manner.
% during the rewrite search.

% The runtime of \textsc{bfr} may also be too expensive for systems 
% where the execution time of most queries is not more than a few tens of 
% seconds.
While this runtime is not trivial, we note that these are complex queries
involving UDFs that run for thousands of seconds.
%Note that these are complex queries involving UDFs that run for thousands
%of seconds.
The amount of time spent to rewrite a query plus the execution
time of the rewritten query is far less than the execution time of the 
original query. %unoptimized 
For instance, Figure~\ref{fig-ue-novf-vs-vf-running-time}(a) 
reports a query execution time of 451 seconds for $A_5$ optimized versus 
2134 seconds for unoptimized. 
Even if the rewrite time for $A_5$ were 1000 seconds 
(it is actually 3.1 seconds here as seen in 
Figure~\ref{fig-ue-bfs-vs-dp-candidates-explored}(c)),
the total execution time would still be 32\% faster than the original query.

% SJ: Removed these as per Carey's comments that it is too late to say these 
% things
% However, additional pruning techniques~\cite{Agra2000} can 
% be used to reduce the number of views, thereby reducing 
% the algorithm search space.
% We note that \WF{} will find the minimum cost rewrite in the space it 
% considers.

\subsubsection{Comparison with Caching-based methods}
\label{sec-experiments-restore}

Next we provide a brief comparison of our approach with caching-based methods 
(such as~\cite{Elgh2012}) that perform only syntactic matching when reusing 
previous results.
With this class of solutions, results can only be reused to answer a new query when 
their respective execution plans are identical, i.e., the query plan and the 
plan that produced the previous results must be syntactically identical.
This means that 
% while two plans may produce identical results, 
if the respective plans differ in any way 
(e.g., different join orders or predicate push-downs), then reuse is not possible.
For instance, with syntactic matching, a query that applies two filters in sequence
$a,b$ will not match a view (i.e., a previous result) 
that has applied the same two filters in a different sequence $b,a$.
In contrast, our {\sc bfr} approach performs semantic matching and query 
rewriting.
In this case, not only will {\sc bfr} match $a,b$ with $b,a$, but it would
also match the query to a view that only has $b$,
by applying an additional filter $a$ during the rewrite process.
% that can only reuse 
%identical answers (i.e., existing views that require no compensations).
% To mimic a syntactic matching method, we create a conservative variant of our 
%  approach that only performs a rewrite if a view and a query have 
% identical plans.  

To represent the class of syntactic caching methods,
we present a conservative variant of our 
approach that performs a rewrite only if a view and a query
have identical $A,F,K$ properties as well as have identical plans.
We term this variant {\sc bfr-syntactic}.
%whereas our approach represents a semantic matching that can perform 
%query rewriting.
%i.e., answers generated by identical plans.

Figure~\ref{fig-restore-comparison-execution-improvement} highlights the 
limitations of caching-based methods by repeating the query evolution 
experiment for Analyst 1 ($A_1v_1$--$A_1v_4$).
We first execute query $A_1v_1$ to produce opportunistic views, and then we
apply both {\sc bfr} and {\sc bfr-syntactic} to queries $A_1v_2$, $A_1v_3$ and 
$A_1v_4$ and report the results in terms of query execution time improvement 
of the solutions produced by {\sc bfr} and {\sc bfr-syntactic}.
Figure~\ref{fig-restore-comparison-execution-improvement} shows that both 
{\sc bfr} and {\sc bfr-syntactic} result in the same execution time 
improvement for $A_1v_2$.
This is because both methods were able to reuse some of the 
(syntactically identical)  views from the previous query. %  ($A_1v_1$).
However, {\sc bfr-syntactic} performs worse than {\sc bfr} for query $A_1v_3$ 
and $A_1v_4$.
This is because {\sc bfr-syntactic} was unable to find many views that were 
exact syntactic matches, whereas {\sc bfr} was able to exploit additional 
views due to {\sc bfr}'s ability to reuse and re-purpose previous results 
through semantic query rewriting.
Even though this particular result is workload dependent, this example 
highlights the fact that while reusing identical results is clearly beneficial,
our approach completely subsumes those that only reuse syntactically identical 
results: even when there are no identical views our method may still produce a 
low-cost rewrite. 

As an additional point of comparison with caching-based methods, we next 
perform an experiment in which we remove \textit{all} identical views from 
the system and then apply our {\sc bfr} algorithm.

% through rewriting.
%While reusing cached results is clearly effective, by only reusing 
% identically generated results, 
%However, performance is highly dependent upon the presence of identical 
%answers previously generated in the system as cache methods lack
%a robust capability to reuse previous results.
\vspace{-0.1in}
\onefig{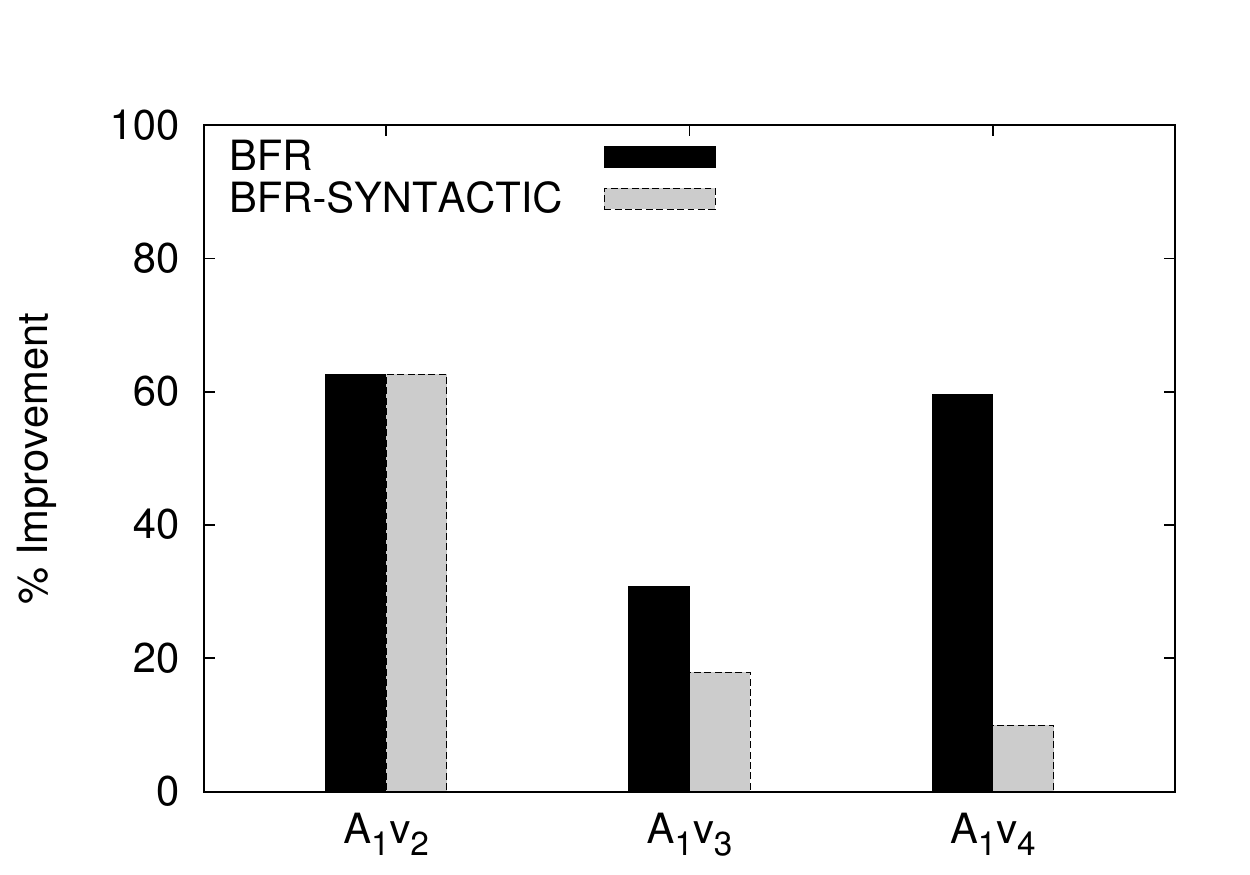}
{Execution time improvement for \textsc{bfr} and \textsc{bfr-syntactic}.}{2in}

\begin{table}[h!] \vspace{-0.25in}
\caption{\label{fig-re-novf-vs-vf-time-improvement}
Execution time improvement without identical views.}
{\small
\centering
\begin{tabular}{| p{1.5cm} | p{0.40cm}| p{0.40cm} | p{0.40cm}| p{0.40cm} 
| p{0.40cm}|p{0.40cm} | p{0.40cm}| p{0.40cm} | p{0.40cm}|}
\hline
Analyst & $A_1$ & $A_2$ & $A_3$ & $A_4$ & $A_5$ & $A_6$ & $A_7$ 
& $A_8$ \\ \hline
% %BFR(with) & 57\% & 64\% & 86\% & 85\% & 78\% & 98\% & 88\% & 84\% \\ \hline
$\WF{}$ & 57\% & 64\% & 83\% & 85\% & 51\% & 96\% & 88\% & 84\% \\ \hline
\end{tabular}
}
\vspace{-0.1in}
\end{table}

% As an additional comparison point of our approach with caching-based methods,
Here we repeat the user evolution experiment after discarding from the 
system all views that are identical to a target in each of the holdout queries 
($A_{1-8}v_1$).
Without these views, syntactic caching-based methods will not be able to find 
\textit{any} rewrites, resulting in 0\% improvement. 
Table~\ref{fig-re-novf-vs-vf-time-improvement} reports the percentage 
improvement for each analyst $A_1$--$A_8$ after discarding all identical 
views. 
This shows {\sc bfr} continues to reduce query execution time dramatically 
even when there are no views in the system that are an exact match to the new 
queries.
The performance improvements are comparable to the result in 
Figure~\ref{fig-ue-novf-vs-vf-running-time}(c) which represents the same 
experiment without discarding the identical views.
Notably there is a drop for $A_5$ compared to the results reported in 
Figure~\ref{fig-ue-novf-vs-vf-running-time}(c) for $A_5$. 
This is because previously in Figure~\ref{fig-ue-novf-vs-vf-running-time}(c), 
$A_5$ had benefited from an identical view corresponding to a 
restaurant-similarity computation that it now has to recompute.
The identical views discarded constituted only 7\% of  the view storage space 
in this experiment, indicating there are many \textit{other} useful views.
Given that analysts pose different but related queries in an evolutionary 
analytical scenario, any method that relies solely on identical matching may 
have limited benefit.

\subsubsection{Storage Reclamation}\label{sec-experiments-storage-reclamation}

\begin{table}
\caption{\label{tab-ga-improvement-vs-storage-reclaimed}
Execution Time Improvement after reclaiming 10\% to 90\% of the 
view storage space.}
{\small
\centering
\begin{tabular}{|c|c|c|c|c|c|c|}
\hline
View Storage Space & $2.0\times$ & $1.8\times$ & $1.5\times$ & $1.0\times$ 
& $0.6\times$ &$0.2\times$\\ \hline
Improvement & 89\% & 89\% & 86\% & 82\% & 58\% & 30\% \\ \hline
\end{tabular}
}
\vspace{-0.1in}
\end{table}

Since storage is not infinite, a reclamation policy is necessary.
Although choosing a beneficial set of views to retain within a finite
storage budget is an interesting problem for future work, here we apply two 
simple policies for storage reclamation.
Our aim in this experiment is to show the robustness of our approach even when
the reclamation policy makes unsophisticated choices such as dropping views 
randomly  %without taking into account their benefit for the workload 
or dropping all of the views identical to a given query.

First, we repeat the experiment in Section~\ref{sec-experiments-ue} using the 
$A_5v_3$ query, but reduce the view storage budget each time.
Retaining all views resulted in a storage space of approximately 
$2.0\times$ the base data size ($\approx$2TB).
The relatively small total size of the views with respect to the log 
base data is due to several reasons.
First, the logs are very wide, as they record a large number of attributes.
However, a typical query only consumes a small fraction of these log 
attributes, which consistent with observations in big data systems.
Second, it is not uncommon for the log attributes to have missing values, 
since the data may be dirty or incomplete.
For instance, in the Twitter log, a tweet may have missing location values. 
Thus a query may discard those tweets without location values.
Third, in this experiment, views are not duplicated since the rewriter makes 
use of existing views whenever possible.  
For these reasons, the total size of the views is relatively small compared 
to the size of the base data.

Table~\ref{tab-ga-improvement-vs-storage-reclaimed} reports the execution 
time improvement for the rewritten query compared to the original query for 
each storage budget.
We repeat each experiment twice, each time randomly dropping views from the 
full set of views, and report the average improvement.
The results show that our method is able to find good rewrites using the 
remaining views available, until the view storage budget is very small.
Second, by the results shown earlier in 
Table~\ref{fig-re-novf-vs-vf-time-improvement}, our method is able to find
good rewrites even when there are no identical views in the system; which 
could be the case if a poor %an adversarial 
 reclamation policy were used.
Designing a good storage reclamation policy is equivalent to the view 
selection problem~\cite{Mami2012} with a storage constraint.

\section{Related Work}

\textbf{\textit{Query Rewriting Using Views.}}
There is a rich body of previous work on rewriting queries using views, but 
these only consider a restricted class of queries.
Representative work includes the popular algorithm MiniCon~\cite{Pott2001}, 
recent work~\cite{Konst2011, Konst2012} showing how rewriting can 
be scaled to a large number of views, and rewriting methods implemented in 
commercial databases~\cite{Gold2001, Zaha2000}.
However, in these works both the queries and views are restricted to the class
of conjunctive queries (SPJ) or additionally include groupby and aggregation 
(SPJGA).

Our work differs in the following two ways:
(a) We show how UDFs can be included in the rewrite process using our UDF model, 
which results in a unique variant of the rewrite problem when there is a 
divergence between the expressivity of the queries and that of the rewrite 
process;
(b) Our rewrite search process is cost-based---\ocost{} enables the 
enumeration of candidate views based on their ability to produce a 
low-cost rewrite.
In contrast, traditional approaches (e.g.,~\cite{Gold2001, Pott2001}) 
typically determine containment first (i.e., if a view can answer a query) 
and then apply cost-based pruning in a heuristic way.
This unique combination of features has not been addressed in the literature
for the rewrite problem.

\textbf{\textit{Online Physical Design Tuning.}}
Methods such as \cite{Schn2007} adapt the physical configuration to benefit a dynamically 
changing workload by actively creating or dropping indexes/views.
Our work is opportunistic, and simply relies on the by-products of query execution
that are almost free.
%During query rewriting, our \WF{} algorithm produces exactly 
%those views required for the optimal rewrite of a workflow
%by exploring the space of existing views as well as new candidate views.
However, view selection methods could be applicable during storage reclamation 
to retain only those views that provide maximum benefit. 
% within a given space budget constraint.

\begin{comment}
\textbf{\textit{View Selection.}}
The view selection problem addresses how to choose views that provide a good
benefit for a representative workload.
Benefit is computed by the reduction in workload cost minus the cost to maintain the views.
In this work we are concerned with the view rewriting problem which determines how
to use any of the existing views to benefit a query.
View maintenance and updates impact the view selection problem rather than
the view rewriting problem we address in this work.
%Although in our system there are no updates-in-place, views
%may need to be maintained as new log data arrives.
\end{comment}

\textbf{\textit{Reusing Computations in MapReduce.}}
Other methods for optimizing MapReduce jobs have been introduced 
such as those that support incremental computations \cite{Li2011},
sharing computation or scans \cite{Nyki2010}, 
and re-using previous results \cite{Elgh2012}.  
% These approaches operate on lower-level constructs 
% (e.g.,plans) 
% than our system.  
As shown in Section~\ref{sec-experiments-restore}, our approach
completely subsumes these methods.
%caching-style sharing methods 
%are effective but provide limited benefit compared to our approach.

\begin{comment}
\textbf{\textit{Evolutionary queries.}}
% These types of queries do not appear to be well-captured by current benchmarks.
Two similar ideas are the interactive OLAP queries included in the TPC-DS~\cite{tpcds}
benchmark and \emph{exploratory} queries in \cite{Khou2009}.
\cite{Khou2009} notes that many users would like to pose 
related queries on a shared data set and subsequently revise and them.
The interactive query class in TPC-DS represents a scenario with
ad-hoc analytic queries designed to mimic
a user analyzing a result before posing a new related query.
%refers to continuously changing queries over 
%log data in an environment outside of the traditional database setting
\end{comment}

\textbf{\textit{Multi-query optimization (MQO).}}
The goal of MQO \cite{Sell1988} (and similar approaches \cite{Nyki2010}) 
is to maximize resource sharing, in particular common intermediate data, by 
producing a scheduling strategy for a set of  in-flight queries.  
%Thus the output  of MQO is a scheduling strategy that minimizes the amount 
%of duplicate work for the query set.
%Our goal is to rewrite a workflow to take advantage of the available 
% materialized views, 
%thus the output of our work is a low-cost rewrite rather than a scheduling 
% policy for multiple queries.
Our work produces a low-cost rewrite rather than a schedule for concurrent 
query plans.
% Our workflow compiler (Section \ref{sec-system-architecture}) does parallelize the nodes
% In the workflow to the extent possible.
% An MQO strategy could be applied to improve the performance of
% parallel nodes if they contain redundant work.
 
\begin{comment} 
\textbf{UDFs in data flows.}
The work in \cite{Hues2012} addresses data flows that contain arbitrary UDFs, the case we
consider as well.  
A static code analysis is applied to the UDF to extract its read set and write set, along with 
its execution signature (M/R jobs) which they use to determine conflicts that affect reordering
UDFs with a fixed signature.
In our work the \textit{outputs} of a UDF have a unique signature, which is then used to perform
semantic matching for rewriting using views.  
Qo-X\cite{Simi2009} also performs optimization of data flows with UDFs for different objectives
(e.g.,cost, recoverability,etc.).  
All of these approaches address plan-level optimizations by considering reordering 
flows that contain UDFs.
Our approach is not concerned with reordering a UDF but rather finding a rewrite among the 
available views that produces the same output of the UDF at a lower cost.
Both \cite{Hues2012} and \cite{Elgh2012} consider an execution path unique if it has the same input files and M/R plan.
A unique execution path is not required for our semantic representation and signature.
\end{comment}

\section{Conclusion}\label{sec-conclusion}

Big data analytics is characterized by exploratory queries with frequent use of 
UDFs containing arbitrary user code. 
To exploit previous results effectively, some understanding of UDFs is required.
In this work, we presented a gray-box UDF model that is simple but 
expressive enough to capture a large class of big data UDFs, enabling our system 
to exploit prior computation.

However, considering many UDFs can make the space of rewrites 
impractical since any UDF in the system may be included in the rewrite process.
We presented a rewrite algorithm \WF{} that utilizes a lower-bound on the cost
of a potential rewrite in order to incrementally grow and search the space of rewrites.
Furthermore, we prove \WF{} is work-efficient and finds the optimal rewrite.  
Our experiments show that for a workload of queries that make extensive use of 
UDFs, our method results in dramatic performance improvements 
with an average of 61\% and up to an order of magnitude.


\begin{thebibliography}{10}

\bibitem{Abit1998}
S.~Abiteboul and O.~M. Duschka.
\newblock Complexity of answering queries using materialized views.
\newblock In {\em PODS}, 1998.

\bibitem{Agra2000}
S.~Agrawal, S.~Chaudhuri, and V.~Narasayya.
\newblock Automated selection of materialized views and indexes in {SQL}
  databases.
\newblock In {\em VLDB}, 2000.

\bibitem{Lefe2013}
{Anonymous PDF document}.
\newblock \url{https://drive.google.com/file/d/0Bzw_KGVFEyOpMFlmUzJEaHB5NkU}.

\bibitem{datafu}
{DataFu}.
\newblock {http://data.linkedin.com/opensource/datafu}.

\bibitem{Elgh2012}
I.~Elghandour and A.~Aboulnaga.
\newblock {ReStore: reusing results of MapReduce jobs}.
\newblock {\em PVLDB}, 5(6), 2012.

\bibitem{Gold2001}
J.~Goldstein and P.-A. Larson.
\newblock Optimizing queries using materialized views: {A} practical, scalable
  solution.
\newblock In {\em SIGMOD}, 2001.

\bibitem{Grum2003}
S.~Grumbach and L.~Tininini.
\newblock On the content of materialized aggregate views.
\newblock In {\em PODS}, 2000.

\bibitem{Hale2001}
A.~Halevy.
\newblock Answering queries using views: A survey.
\newblock {\em VLDBJ}, 10(4), 2001.

\bibitem{Hjal03}
G.~Hjaltason and H.~Samet.
\newblock Index-driven similarity search in metric spaces.
\newblock {\em TODS}, 28(4), 2003.

\bibitem{Hues2012}
F.~Hueske, M.~Peters, M.~J. Sax, A.~Rheinl\"{a}nder, R.~Bergmann, A.~Krettek,
  and K.~Tzoumas.
\newblock Opening the black boxes in data flow optimization.
\newblock {\em PVLDB}, 5(11), 2012.

\bibitem{Jayr2006}
T.~S. Jayram, P.~G. Kolaitis, and E.~Vee.
\newblock The containment problem for {Real} conjunctive queries with
  inequalities.
\newblock In {\em PODS}, 2006.

\bibitem{Konst2012}
G.~Konstantinidis and J.~Ambite.
\newblock Optimizing query rewriting for multiple queries.
\newblock In {\em IIWeb}, 2012.

\bibitem{Konst2011}
G.~Konstantinidis and J.~L. Ambite.
\newblock Scalable query rewriting: a graph-based approach.
\newblock In {\em SIGMOD}, 2011.

\bibitem{lefe2013b}
J.~Le{F}evre, J.~Sankaranarayanan, H.~Hac{\i}g\"{u}m\"{u}\c{s}, J.~Tatemura,
  and N.~Polyzotis.
\newblock Towards a workload for evolutionary analytics.
\newblock In {\em {SIGMOD Workshop on Data Analytics in the Cloud (DanaC)}},
  2013.

\bibitem{Levy1995}
A.~Y. Levy, A.~O. Mendelzon, and Y.~Sagiv.
\newblock Answering queries using views (extended abstract).
\newblock In {\em PODS}, 1995.

\bibitem{Li2011}
B.~Li, E.~Mazur, Y.~Diao, A.~Mc{G}regor, and P.~Shenoy.
\newblock {A platform for scalable one-pass analytics using MapReduce}.
\newblock In {\em SIGMOD}, 2011.

\bibitem{Mami2012}
I.~Mami and Z.~Bellahsene.
\newblock A survey of view selection methods.
\newblock {\em SIGMOD Rec.}, 41(1), 2012.

\bibitem{Nyki2010}
T.~Nykiel, M.~Potamias, C.~Mishra, G.~Kollios, and N.~Koudas.
\newblock {MRShare: sharing across multiple queries in MapReduce}.
\newblock {\em VLDB}, 3(1-2), 2010.

\bibitem{piggybank}
{PiggyBank}.
\newblock {https://cwiki.apache.org/confluence/display/PIG/PiggyBank}.

\bibitem{Pott2001}
R.~Pottinger and A.~Halevy.
\newblock {MiniCon}: A scalable algorithm for answering queries using views.
\newblock {\em VLDB}, 10(2), 2001.

\bibitem{Schn2007}
K.~Schnaitter, S.~Abiteboul, T.~Milo, and N.~Polyzotis.
\newblock On-line index selection for shifting workloads.
\newblock In {\em ICDE}, 2007.

\bibitem{Sell1988}
T.~Sellis.
\newblock Multiple-query optimization.
\newblock {\em {TODS}}, 13(1), 1988.

\bibitem{Zaha2000}
M.~Zaharioudakis, R.~Cochrane, G.~Lapis, H.~Pirahesh, and M.~Urata.
\newblock Answering complex {SQL} queries using automatic summary tables.
\newblock In {\em SIGMOD}, 2000.

\end{thebibliography}
\end{document}